\documentclass[prd,preprint,nofootinbib]{revtex4}

\usepackage{amsmath,amssymb}
\usepackage{dcolumn}% Align table columns on decimal point
\usepackage{bm}% bold math
\usepackage{graphicx,psfrag}

\newcommand{\dslash}{\ensuremath{\displaystyle{\not}}}

\begin{document}
\title{Realistic Composite Higgs Models}

\date{\today}

\author{Charalampos Anastasiou}
\author{Elisabetta Furlan}
\author{Jos\'e Santiago}
\affiliation{Institute for Theoretical Physics, ETH, CH-8093,
  Zurich, Switzerland}

\begin{abstract}
We study the role of fermionic resonances in realistic
composite Higgs models. We consider the low energy effective
description of a model in which the Higgs arises
as the pseudo-Goldstone boson of an $SO(5)/SO(4)$ global symmetry
breaking pattern. Assuming that only fermionic resonances are present
below the cut-off of our effective theory, we perform a detailed
analysis of the electroweak constraints on such a model. This 
includes the exact one-loop calculation of the $T$ parameter and the anomalous
$Z b_L\bar{b}_L$ coupling for arbitrary new fermions and couplings. 
Other relevant observables, like $b\to s \gamma$ and $\Delta B=2$
processes have also been examined. We find that,
while minimal models are difficult to make compatible with electroweak
precision tests, models with several fermionic resonances, such as  the
ones that appear in the spectrum of viable composite Higgs models from warped extra
dimensions, are fully realistic in a large region of parameter space. These fermionic
resonances could be the first observable signature of the model at the LHC.
\end{abstract}


\maketitle

\section{Introduction}
\label{intro}
One main objective of the Large Hadron Collider (LHC) 
is to discover the precise mechanism of electroweak symmetry 
breaking (EWSB). 
A well motivated hypothesis is that  there exists a Higgs boson which is not 
a fundamental scalar. Instead it could be a composite state of a strongly coupled theory,  
the pseudo-Goldstone boson  of a spontaneously broken global symmetry~\cite{Kaplan:1983fs}.  
Compositeness can explain the insensitivity of EWSB to ultraviolet physics,  
while the pseudo-Goldstone nature of the Higgs boson  may explain the little 
hierarchy between the scale of new physics and the scale of EWSB. 

This mechanism has recently received increased
attention, due to the realization that calculable composite Higgs
models can be constructed in
five dimensions~\cite{Contino:2003ve}. The main idea is an old 
one~\cite{Manton:1979kb}, but only recently realistic models
in warped extra dimensions~\cite{Contino:2003ve,Agashe:2004rs,Agashe:2005dk,Contino:2006qr,Panico:2008bx}
have been constructed. 
The experience with five-dimensional models indicates that custodial
symmetry~\cite{Agashe:2003zs} and a custodial protection of the $Zb_L
\bar{b}_L$ coupling~\cite{Agashe:2006at} are likely ingredients of
realistic constructions.
Little Higgs models~\cite{ArkaniHamed:2001nc} also use the idea of a
Higgs boson which is the pseudo-Goldstone boson of a spontaneously broken 
global symmetry. In order to solve the little hierarchy problem, 
they employ the mechanism of collective symmetry
breaking, which ensures that the Higgs mass remains insensitive to
ultraviolet physics at one loop. 

The main phenomenological implications of a Higgs boson which is  the
pseudo-Goldstone boson of an extended broken symmetry are largely independent 
of the particular details of how the global symmetry is broken. They can 
therefore be conveniently described using an effective Lagrangian 
approach~\cite{Barbieri:2007bh,Giudice:2007fh}. A reasonable starting point is a symmetry breaking pattern that
includes custodial symmetry. A minimal example of such a pattern is
given by a global $SO(5)$ symmetry broken at a scale $f$ to its
custodially symmetric subgroup $SO(4)$~\cite{Agashe:2004rs,Agashe:2005dk}.

Using this effective Lagrangian approach, it was argued
in~\cite{Barbieri:2007bh} that composite Higgs models with
an $SO(5)/SO(4)$ symmetry breaking pattern   
are difficult to make compatible with electroweak and flavor precision
data without introducing a substantial fine-tuning. 
The argument was based on very minimal models, in which
fermionic resonances did not span full representations of the $SO(5)$
group.  In addition,  the estimation of electroweak observables was made 
neglecting contributions which are formally subleading, but can be relevant in specific situations.
Two recent works extended the analysis of~\cite{Barbieri:2007bh}  
to models in which the fermionic composites span full representations of
$SO(5)$~\cite{Lodone:2008yy,Gillioz:2008hs}. Although these analyses
differ in several aspects, like the degree of explicit $SO(5)$
symmetry breaking and the symmetry breaking patterns, their outcome is somewhat similar.  Only  
a small region of parameter space is allowed by electroweak
precision data in models with no significant fine-tuning and one set of
fermionic composites spanning a vector representation of $SO(5)$.

In this paper, we investigate thoroughly the viability of models with an $SO(5)/SO(4)$  
symmetry breaking pattern. Our analysis extends previous works in two ways, 
by making a careful computation of the  effect of the new fermionic 
states on electroweak observables and by considering the effect of an extended fermionic sector. 

In our study of electroweak precision constraints we use an exact one-loop calculation of the relevant electroweak observables. 
We do this in complete generality, and  our  analytic formulae 
can be used in other models. In particular, our result for the
anomalous $Z b_L \bar{b}_L$ coupling is, to the best of our knowledge, 
the first complete calculation for an arbitrary number of new 
quarks with generic couplings. We find that such 
an exact computation  can be important when formally subleading effects in commonly employed  approximations are enhanced.

We also examine the possibility of multiple sets of fermionic composites, 
departing from minimal constructions. Our motivation is the model presented  
in~\cite{Panico:2008bx}, in which a five-dimensional realization of a composite Higgs model 
with $SO(5)/SO(4)$ symmetry breaking pattern was shown to be fully
compatible with electroweak precision tests, flavor observables, electroweak symmetry 
breaking and the observed dark matter relic abundance. The non-minimal fermion
sector of the model in~\cite{Panico:2008bx}   was indispensable in order to 
render its predictions  compatible with experimental data. 
The goal of our article is to present an effective four-dimensional 
description of composite Higgs models with a non-minimal fermionic content and discuss their 
electroweak and flavor constraints. We show that there are large regions 
of parameter space in which composite Higgs models can provide a fully 
realistic description of EWSB without fine tuning.

The paper is organized as follows. In section~\ref{effective} 
we briefly review the
effective description of composite Higgs models with an $SO(5)/SO(4)$
symmetry breaking pattern, including the experimental constraints of
minimal models. Section~\ref{fermions} is devoted to a description of the
relevant fermionic sector of the theory and its effects on electroweak
precision observables and flavor physics. 
In section~\ref{phenomenology} we  present exact one-loop expressions for  
electroweak and flavor  precision observables which are valid in general extensions of the SM. 
The main phenomenological implications of our model are discussed in section~\ref{ewpt}, where 
we describe two options for realistic composite Higgs models, a very simple one
and a slightly more involved one which is closer to the realistic examples we
know from extra dimensions.
Finally, we conclude in section~\ref{sec:conclusions}. 

%%%%%%%%%%%%%%%%%%%%%%%%%%%%%%%%%%%%%%%%%%%%%%%%%%%%%%%

\section{Effective Description of Composite Higgs
  Models}
\label{effective} 
The low energy effective description of a composite Higgs model with
$SO(5)/SO(4)$ symmetry breaking pattern can be described
by a scalar $\phi$ in the fundamental representation of $SO(5)$,
subject to the constraint
\begin{equation}
\phi^2=f^2, \label{f:definition}
\end{equation}
where $f$ is the scale of the global symmetry breaking, assumed to be
somewhat larger than the EWSB scale $v\approx 174$ GeV. The first four
components of $\phi$, which transform as the fundamental representation
of SO(4), are denoted by $\vec{\phi}$. The $SU(2)_L\times U(1)_Y$ subgroup 
of $SO(4)= SU(2)_L \times SU(2)_R$,
where the hypercharge corresponds to the $T^3_R$ generator, is weakly
gauged.~\footnote{An extra $U(1)$ group is required to generate 
 the correct Weinberg angle, but it is irrelevant for the present discussion
 and will be disregarded.} The vacuum expectation value
of $\vec{\phi}$ breaks the EW symmetry,
\begin{equation}
m_W^2=\frac{g^2 v^2}{2}, \quad v^2=\frac{1}{2}\langle \vec{\phi}^2
\rangle.
\end{equation}
From Eq.~(\ref{f:definition}) we see that the ratio
\begin{equation}
s_\alpha \equiv \sin\alpha\equiv \sqrt{2}\frac{v}{f},
\end{equation}
measures the Higgs compositeness, \textit{i.e.} 
how the vev of $\phi$ is split between $\vec{\phi}$ and
$\phi_5$. Canonical normalization of the different components in
$\phi$, expanded around its vev, requires a rescaling of the physical
Higgs 
\begin{equation}
h \to \cos \alpha~ h \equiv c_\alpha~ h,
\end{equation}
whereas the would be Goldstone bosons are not
modified. This redefinition implies an important feature of Higgs
compositeness, namely that Higgs
couplings to gauge bosons 
are suppressed with respect to the couplings
in the SM by the factor $c_\alpha=\sqrt{1-2v^2/f^2}$. In
  the case of fermions, the suppression factor depends also on the
  embedding of the SM fermions in $SO(5)$
  representations~\cite{Contino:2006qr,Falkowski:2007hz}.
This suppression of the Higgs couplings affects the quantum corrections to
electroweak precision observables, leading to some sensitivity to the
ultraviolet cut-off~\cite{Barbieri:2007bh}. 
The leading effect can be taken into account by replacing the Higgs mass
with an effective (heavier) Higgs in the SM expressions of the one-loop 
corrections to the electroweak precision observables,
\begin{equation}
m_\mathrm{EWPT,eff}=m_h(\Lambda/m_h)^{s^2_\alpha},
\end{equation}
where $\Lambda = 4 \pi f / \sqrt{N_G} = 2 \pi f$ is the ultraviolet
cut-off of our effective theory, $N_G$ being the number of Goldstone bosons. This 
modification gives rise to an additional contribution to the
Peskin-Takeuchi~\cite{Peskin:1991sw} $S$ and $T$ parameters
\begin{equation}
\Delta S = \frac{1}{12\pi}\ln \left(
\frac{m_\mathrm{EWPT,eff}^2}{m_\mathrm{h,ref}^2}\right),
\quad
\Delta T = -\frac{3}{16\pi c_W^2}\ln \left(
\frac{m_\mathrm{EWPT,eff}^2}{m_\mathrm{h,ref}^2}\right), \label{ST:Higgs}
\end{equation}
where $m_\mathrm{h,ref}$ is the reference Higgs mass used in the
electroweak fit and $c_W$ is the cosine of the Weinberg angle.

Furthermore, custodial symmetry can naturally account for a suppressed
contribution of ultraviolet physics to the $T$ parameter, but there is
no reason not to expect a contribution to the $S$ parameter from
higher dimensional operators. A reasonable estimate, assuming that
new physics couples linearly to the SM (otherwise this estimate should
have an extra loop suppression), is
\begin{equation}
\Delta S_\Lambda \sim \frac{4 s_W^2}{\alpha_{em}} \frac{g^2 v^2}{\Lambda^2}
\approx 0.16 \left(\frac{3\mbox{ TeV}}{\Lambda}\right)^2. \label{S:Lambda}
\end{equation}

\begin{figure}[t,b]
\includegraphics[width=.5\textwidth]{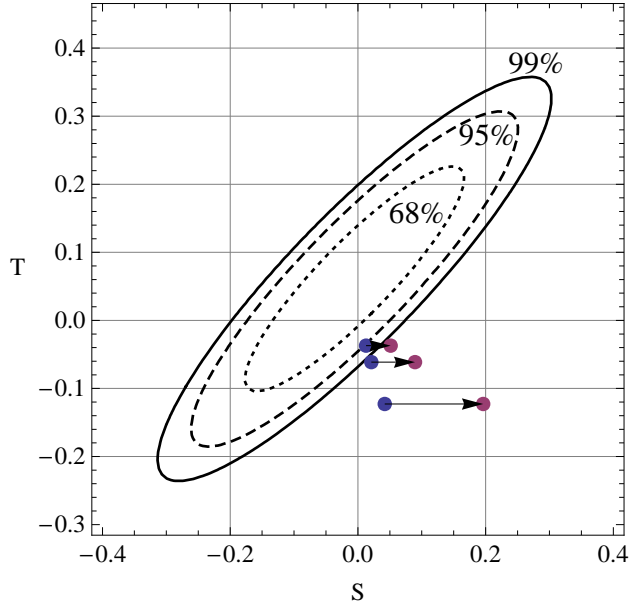}
\caption{ 
$68\%$, $95\%$ and $99\%$ C.L. limits on $S$ and $T$ for 
a fit to electroweak observables with three independent 
parameters ($S$, $T$ and $\delta g_{b_L}$),
fixing the optimal value of $\delta g_{b_L}=-2.5\times 10^{-4}$ 
and a reference Higgs mass
$m_{\mathrm{h,ref}}=120$ GeV. 
The dots at the start and the tip of the arrows show the effects due to Higgs compositeness, 
Eq.~(\ref{ST:Higgs}), and to UV effects on $S$, Eq.~(\ref{S:Lambda}),
respectively. The effect is shown (from top to bottom) 
for the values of $s_\alpha = 0.25, 0.33$ and $0.5$.} 
\label{STplot} 
\end{figure}

The combination of the two corrections, (\ref{ST:Higgs}) and
(\ref{S:Lambda}), results in a positive shift to the $S$ parameter and
a negative shift to the $T$ parameter. We show in Fig.~\ref{STplot}
the current constraints on the $S$ and $T$ parameters, assuming a
Higgs mass $m_h=120$ GeV. We also show the contributions from Higgs
compositeness and UV physics for different values of $s_\alpha$.
These constraints are obtained as follows. For
reasons that will become apparent in the next section, we have
performed a fit to all the relevant electroweak observables, allowing
the $S$ and $T$ parameters and an anomalous coupling of the $b_L$
quark to the $Z$, that we denote by $\delta g_{b_L}$, 
to vary (note that we have set $U=0$ in our fit, as it is expected to
be vanishingly small in our model).~\footnote{We use Ref.~\cite{Han:2004az} 
for the fit updated to the most recent experimental 
data~\cite{Collaboration:2008ub}. $\alpha_s(m_Z)=0.1183$
and $m_t=172.4$ GeV have been fixed to the best fit values from a 5
parameter ($\alpha_s(m_Z)$,
$m_t$, $S$, $T$ and $\delta g_{b_L}$). We would like to thank 
E. Pont\'on for help with the fit and with Fig.~\ref{STplot}.}
In Fig.~\ref{STplot}, we have fixed the optimal value of $\delta
g_{b_L}=-2.5\times 10^{-4}$, although the result of a proper 
projection over the $S-T$ plane does not differ significantly.
As we see in the figure, the above corrections put the model in gross
contradiction with experimental observations for any sizable Higgs
compositeness. As emphasized 
in~\cite{Barbieri:2007bh}, this is a quite generic implication of
Higgs compositeness and custodial symmetry, that seems to disfavor
composite Higgs models as a realistic description of electroweak
symmetry breaking. However, we have not included yet in our discussion the
effect of other composite states that might be lighter than the
cut-off of our effective description. In the next section we discuss
the effect that new fermionic states can have on electroweak precision
tests.~\footnote{New bosonic resonances have been recently shown to
be able to make even Higgs--less models compatible with experimental
data~\cite{Barbieri:2008cc}.}  

%%%%%%%%%%%%%%%%%%%%%%%%%%%%%%%%%%%%%%%%%%%%%%%%%%%%%%%

\section{The Fermionic Sector\label{fermions}}
\label{models}
If the Higgs is a composite state of a new strongly interacting
theory, it is natural to assume that the large top mass is due to
partial top compositeness. The partners under $SO(5)$ of the composite
states with which the top mixes can then play an important role in
electroweak precision tests. 
In particular, if the top sector is
partly composite, large corrections to the $Z b_L \bar{b}_L$ coupling
are typically expected, unless some symmetry forbids
them~\cite{Agashe:2006at}. Fermions, in the fundamental or adjoint representation of the $SO(5)$
group, are natural building blocks that incorporate the left-right
symmetry guaranteeing the absence of large tree-level
corrections to the $Z b_L \bar{b}_L$ coupling. In this article
we will consider the former possibility and include new (composite)
vector-like fermions that
transform in the five-dimensional vector representation of $SO(5)$,
which decomposes under $SO(4)=SU(2)_L\times SU(2)_R$ as
\begin{equation}
\Psi = (Q,X,T) \Rightarrow (5)=(2,2)\oplus (1).
\end{equation} 
$Q$ and $X$ form a bidoublet under $SU(2)_L\times SU(2)_R$, they are
$SU(2)_L$ doublets with hypercharges $1/6$ and $7/6$, respectively
(and $T_3^R=-1/2$ and $1/2$, respectively). $T$ is an $SU(2)_L\times
SU(2)_R$ singlet with hypercharge $2/3$. The SM quarks $q_L$ and $t_R$
have the same quantum numbers under the SM gauge group as $Q$ and $T$,
respectively. 

\begin{figure}[H,B,t,b]
\psfrag{DTbyDTt}[t][t][1.]{$\Delta T/\Delta T_{SM}$}
\psfrag{DT}[t][t][1.]{$\Delta T$}
\psfrag{dg}[t][t][1.]{$10^3 \delta g_{b_L}$}
\psfrag{dgbydgt}[t][t][1.]{$\delta g_{b_L}/\delta g_{b_L}^{SM}$}
\includegraphics[width=.5\textwidth]{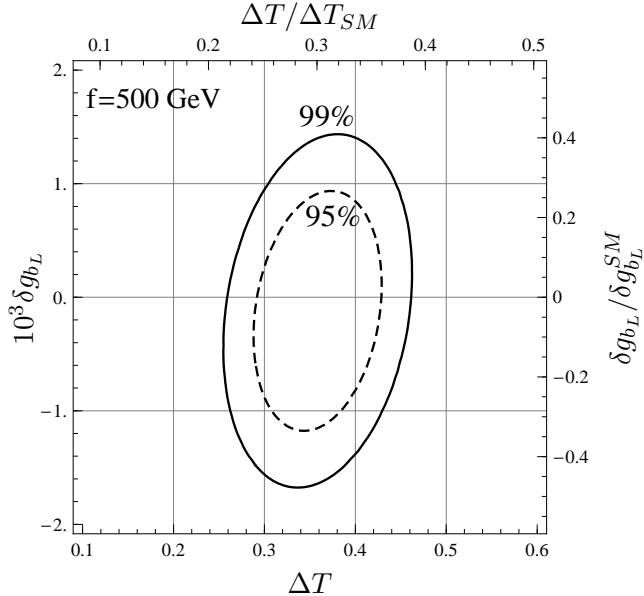} 
\caption{ 
$95\%$ and $99\%$ C.L. limits on $T$ and $\delta g_{b_L}$ for 
a fit to electroweak observables with three independent parameters
($S$, $T$ and $\delta g_{b_L}$) assuming $s_\alpha \approx 0.5$ ($f  = 500$ GeV) 
and a reference Higgs mass $m_{\mathrm{h,ref}}=120$ GeV
(no projection). The top and right axes show the value of the $T$
parameter and $\delta g_{b_L}$ in terms of the SM one-loop contribution. 
For this value of $s_\alpha$ there is no allowed region at $68\%$ C.L..
}
\label{TgbLplot} 
\end{figure}
The question is, given the large corrections to the $S$ and $T$
parameters from the Higgs sector, can the one-loop contribution of
this new fermionic sector to $T$ be such that the model is compatible
with experimental data? As shown in Fig.~\ref{STplot}, a 
large positive contribution to $T$ is required, but this has to be
obtained without spoiling the good agreement with other experimental
observables. In particular, $\delta g_{b_L}$ is extremely constrained
experimentally.
The situation is exemplified in Fig.~\ref{TgbLplot}, in which we show
the constraints on the contribution to the $T$ parameter and
$\delta g_{b_L}$ from the new quarks, for a fixed $s_\alpha \approx  0.5$
($f  = 500$ GeV). In the figure, we have shown both the actual value of
the allowed $T$ and $\delta g_{b_L}$ (bottom and left axes,
respectively) together with the values relative to the one-loop SM
corrections (top and right axes). We see that, for this value of
$s_\alpha$, even at the $99\%$ C.L. the $T$ parameter must  be
positive and it is allowed to vary at most by about $20\%$ in units of the SM one-loop
correction. We also observe that $\delta g_{b_L}$ is constrained in the range 
$-0.5 \lesssim \delta g_{b_L}/\delta g_{b_L}^ {SM} \lesssim 0.4$. 

The effects on the $T$ parameter and the $Z b_L \bar{b}_L$ coupling  
of new vector-like quarks with the above quantum numbers through their
mixing with the top were discussed in detail in
Ref.~\cite{Carena:2006bn}. 
The $SU(2)_L$ quantum numbers of the
different multiplets determine, to a great degree, the sign of the
contribution to the $T$ parameter. Through their mixing with the top,
the different $SU(2)_L$ multiplets typically contribute (assuming only one type
of mixing at a time) as follows: 
\begin{itemize}
\item Singlets ($T$) contribute positively to the $T$ parameter and
  the $Z b_L \bar{b}_L$ coupling.
\item Hypercharge $7/6$ doublets ($X$) contribute negatively to the
  $T$ parameter.
\item Hypercharge $1/6$ doublets ($Q$) contribute positively to the
  $T$ parameter.
\end{itemize}
Furthermore, there is a strong correlation between the contribution of
vector-like singlets to the $T$ parameter and the $Zb_L \bar{b}_L$
coupling (they are both positive and governed by the same parameters)
which implies a large positive correction to the anomalous $Z b_L
\bar{b}_L$ coupling (making it less compatible with experimental
observations) in the case that the singlet contributes a large
positive amount to $T$. Due to the particular chirality of the final
state, doublets ($Q,X$) do not contribute significantly to the $Z b_L
\bar{b}_L$ coupling (they can contribute to the $Z b_R \bar{b}_R$
coupling but that is a more model dependent issue that is not
correlated with the contribution to the $T$ parameter).

Given the constraints shown in Fig.~\ref{TgbLplot}, it is clear that a
large contribution
from the singlet is bound to give problems, as it will also give a
large contribution to the $Zb_L \bar{b}_L$ coupling, which is extremelly
constrained experimentally. Ideally, one would want a large
contribution from $Q$, but this is also difficult, due to the
constraints imposed by the global $SO(5)$ symmetry which usually mean
a larger (negative) contribution from $X$. The general situation is
more complicated, and several modes can give large contributions to
make the model compatible with EWPT. The difficulties we have
outlined, however, mean that the model can be quite predictive, and only
a few patterns can be realized. Our goal is to describe these patterns
and their implications at the LHC.

We restrict our discussion to the top sector~\footnote{For
simplicity we will assume the bottom
mass to come from a direct Yukawa coupling, $\bar{q}_L \phi b_R$. This
small explicit breaking of
the SO(5)/SO(4) pattern will not have appreciable effects.} $q_L, t_R$, which couples to a set of new vector-like quarks,
transforming as the $(5)$ representation of $SO(5)$,
$\Psi_{L,R}=(Q,X,T)_{L,R}$ as
\begin{equation}
-\mathcal{L}_\mathrm{int}=m_L^{i} \bar{q}_L Q^i_R + m_R^{i}
\bar{T}^i_L t_R + \mathrm{h.c.}. \label{L:int}
\end{equation}
The new sector is $SO(5)$ invariant, with a mass Lagrangian given by
\begin{equation}
-\mathcal{L}_\mathrm{SO(5)}= 
m^i_{\Psi}\bar{\Psi}^i \Psi^i
+\frac{\mu_{ij}}{f} (\bar{\Psi}^i \phi) 
(\phi^\mathrm{T} \Psi^j), \label{L:BSM}
\end{equation}
where the indices $i,j$ allow for the possibility of more than one set
of fermionic composites and the brackets in the second term indicate
the contraction of the $SO(5)$ indices. $\mu_{ij}$ is a hermitian matrix.
$\phi$ is our scalar 5-plet and the terms involving it can be
written, in an $SU(2)_L \times U(1)_Y$ invariant way, as
\begin{equation}
\bar{\Psi}\phi= \bar{Q}\tilde{\varphi}+\bar{X} \varphi + \bar{T}
\phi_5, 
\end{equation}
where $\varphi$ and $\tilde{\varphi}=i \sigma^2 \varphi^\ast$ are the
SM Higgs doublets with hypercharge $1/2$ and $-1/2$, respectively.
Note that these have to incorporate the rescaling of the physical
Higgs. For instance we have
\begin{equation}
\varphi = 
\begin{pmatrix}
\varphi^+ \\
v + c_\alpha \frac{\varphi_0+\varphi_0^\ast}{2}
+\frac{\varphi_0-\varphi_0^\ast}{2}
\end{pmatrix},
\end{equation}
with $\langle \varphi_0 \rangle = \langle \varphi^+ \rangle =0$.
We also have $\phi_5=f c_\alpha- \frac{v}{f}(\varphi_0+\varphi_0^\ast)
+ \ldots$, where the dots denote terms with two or more scalars.
From these expressions and the Lagrangians in Eqs.~(\ref{L:int}) and
(\ref{L:BSM}) we can compute the mass matrix and the Yukawa couplings for
the quarks in the model (including the couplings to the would be
Goldstone bosons that will be required for the calculation of the anomalous
$Zb_L \bar{b}_L$ coupling).
The mass terms can be written in matrix form as
\begin{equation}
- \mathcal{L}=\overline{\begin{pmatrix}
t_L \\ Q^u_L \\ X^u_L \\ T_L 
\end{pmatrix}}
\begin{pmatrix}
0 & m_L^\mathrm{T} & 0 & 0 \\
 0 & m_\Psi+\frac{s_\alpha^2}{2}f\mu 
& \frac{s_\alpha^2}{2}f\mu  & c_\alpha v \mu \\
0 & \frac{s_\alpha^2}{2}f\mu  & m_\Psi+\frac{s_\alpha^2}{2}f\mu   &
c_\alpha v \mu \\ 
m_R  & c_\alpha v \mu  & c_\alpha v \mu 
& m_\Psi + c_\alpha^2f \mu 
\end{pmatrix}
\begin{pmatrix}
t_R \\ Q^u_R \\ X^u_R \\ T_R
\end{pmatrix}+ \mathrm{h.c.},
\label{eq:mass-matrixF}
\end{equation}
where we have implicitly written the mass matrix in block form and
$Q^u_{L,R}$, $X^u_{L,R}$ are the charge $2/3$ 
components of $Q$ and $X$, respectively.
The interaction Lagrangian, Eq.~(\ref{L:int}), gives a mixing between
the fundamental fields $q_L$, $t_R$ and the composite states $Q$ and
$T$, which makes the Lagrangian non-diagonal \textit{before} EWSB. If
we diagonalize the mass matrix before EWSB (with $v=0$), we will end
up with a massless $SU(2)_L$ doublet and a massless singlet that are
now an admixture of fundamental and composite states. These new massless
states have Yukawa couplings, thanks to their composite components
(since, assuming that the only explicit breaking of $SO(5)$ is through $m^i_{L,R}$, the Higgs only couples to composites). 

In order to better
understand this mechanism, we consider
the case that there is only one set of composite fermionic states
below the cut-off of our theory. We can then
diagonalize the mass matrix, for $v=0$,
by means of the following rotations
\begin{equation}
q_L \to \cos \theta_L q_L + \sin \theta_L Q_L, \quad
Q_L \to -\sin \theta_L q_L + \cos \theta_L Q_L,
\end{equation}
and
\begin{equation}
t_R \to \cos \theta_R t_R + \sin \theta_R T_R, \quad
T_R \to -\sin \theta_R t_R + \cos \theta_R T_R.
\end{equation}
Note that in the case of $q_L$ and $Q_L$ we are rotating entire
doublets; these have the same quantum numbers and therefore no  traceable physical footprint of the rotation is left.
The mixing angles determining the degree of compositeness of
$q_L$ and $t_R$ are, respectively,
$\tan \theta_L=\frac{m_L}{m_\Psi}$ 
and $\tan \theta_R=\frac{m_R}{m_T}$, where we have defined
$m_T\equiv m_\Psi+f \mu$ (note that now $\mu$
is not a matrix but a number and that for $v=0$ we have $c_\alpha=1$).

With these field rotations, the mass Lagrangian for the charge $2/3$
quarks reads
\begin{equation}
- \mathcal{L}_m=
\overline{
\begin{pmatrix}t_L \\ Q^u_L \\X^u_L\\ T_L
\end{pmatrix}}
\begin{pmatrix}
s_L s_R c_\alpha v \mu & -s_L \frac{s_\alpha^2}{2} f \mu 
& -s_L \frac{s_\alpha^2}{2} f \mu  & -s_L c_R c_\alpha v \mu \\
-s_R c_L c_\alpha v \mu & 
\frac{m_\Psi}{c_L}+c_L \frac{s_\alpha^2}{2} f \mu  
& c_L \frac{s_\alpha^2}{2} f \mu   & c_L c_R c_\alpha v \mu \\
-s_R c_\alpha v \mu & \frac{s_\alpha^2}{2} f \mu  
& m_\Psi +\frac{s_\alpha^2}{2} f \mu 
& c_R c_\alpha v \mu \\
s_R s_\alpha^2 f \mu & c_\alpha v \mu & c_\alpha v \mu & 
\frac{m_T}{c_R}-c_R s_\alpha^2 f \mu
\end{pmatrix}
\begin{pmatrix}
t_R \\ Q^u_R \\ X^u_R \\ T_R
\end{pmatrix}
+ \mathrm{h.c.}
\label{eq:mass-matrix}
\end{equation}
where we have denoted $s_{L,R}\equiv \sin \theta_{L,R}$,
$c_{L,R}\equiv \cos \theta_{L,R}$.

We already see in this mass Lagrangian some of the constraints imposed
by the global symmetry. First, the top quark acquires mass through its
mixing with the composite states. In order to have a large enough top
mass, $t_L$ and $t_R$ cannot be \textit{simultaneously} mostly
fundamental (small $s_L$ and $s_R$). Second, the mixing of the
hypercharge $1/6$ doublet with the top (one possible source of
positive contribution to the $T$ parameter) is always smaller, by a
factor $c_L$, than the mixing of the hypercharge $7/6$ multiplet. If
the two are degenerate or $X$ is lighter than $Q$ (as happens in
minimal five-dimensional models), then the system $Q,X$ usually
contributes negatively to the $T$ parameter.  This effect together with the additional 
negative contribution from the fact that the Higgs boson is composite lead
to the problem discussed previously.  
Either the fermion contribution to the $T$ parameter is
not large enough and therefore incompatible with electroweak precision data
(given the positive contribution to the $S$ parameter from UV physics)
or large corrections to flavor preserving and violating $b$ couplings
are introduced, again in conflict with experimental data. 

A precise assessment of the model viability requires  a study after 
a complete diagonalization of the  matrix in Eq.~(\ref{eq:mass-matrix}), since several modes
can simultaneously give relevant contributions which are difficult to
disentangle qualitatively. A detailed analysis of electroweak constraints in our
model is therefore required to see if there are regions of parameter
space compatible with current data. This detailed analysis includes
a precise calculation of the main electroweak observables, which in
our case can be encoded in the $T$ parameter and $\delta g_{b_L}$,
including formally subleading contributions not proportional to large
Yukawa couplings and a careful scan over parameter space. This is the
subject of the next two sections.

%%%%%%%%%%%%%%%%%%%%%%%%%%%%%%%%%%%%%%%%%%%%%%%%%%%%%%%

\section{Evaluation of precision electroweak observables}
\label{phenomenology}
In this section, we compute the one-loop contribution of the new
fermionic sector to the most relevant electroweak observables, 
which receive large corrections due to the large  value 
of the top mass. The most important observables
are the $T$ parameter and the anomalous $Z b_L \bar{b}_L$
coupling.~\footnote{The one-loop contribution
from the fermionic sector to $S$, which was computed
in~\cite{Carena:2006bn}, is negligible
compared with the tree level contribution from UV physics.} 
Other observables, $B_{d,s}-\bar{B}_{d,s}$ mixing, $B_{d,s}\to
\mu^+\mu^-$, and $b \to s \gamma$, may also receive large one-loop 
corrections which are however less generic, depending for example
on the details of how the bottom quark gets a mass. 
These  observables provide  typically weaker constraints than 
the $T$ parameter and the anomalous $Z b_L \bar{b}_L$
coupling (see for instance~\cite{Haisch:2007ia} 
for a discussion in the context of MFV
scenarios). Nevertheless, we have explicitly checked that the constraints 
from $B-\bar{B}$ mixing and $b\to s \gamma$ are indeed typically weaker in
our model.

Given the stringent constraints on the new fermionic contributions to the 
$T$ parameter  and $\delta g_{b_L}$  we have found  it important to 
calculate  these observables  as  precisely as possible. In our results we do not 
discard any one-loop diagram.  For $\delta g_{b_L}$, in particular, 
we compute the full dependence of  the corresponding amplitude  on all 
mass parameters (including the $Z$, $W$ and Goldstone boson masses) except
for the bottom or lighter quark masses. 
Our calculation of $\delta g_{b_L}$ is  general and the result can be readily 
used in other models.~\footnote{Our calculations are
done assuming point particles. The observables computed here are however
finite and dominated by scales of the order of the masses of the
particles involved. In the composite Higgs models we are interested in, the relevant masses
are much smaller than the strong coupling scale and therefore our approximation
is valid.} 

We perform all our calculations in the 't Hooft-Feynman gauge, in which the
Goldstone bosons and the corresponding gauge bosons have the same
mass. We consider an arbitrary number of new quarks $\psi^i_Q$, with
electric charge $Q$ ($Q=-1/3, 2/3$ or $5/3$ in our model) and mass $m^i_Q$. 
We parametrize their couplings to the $Z$ and $W$ bosons in the mass
eigenstate basis as (an implicit sum over the particle charges $Q$ is always understood)
\begin{eqnarray}
\mathcal{L}^Z&=& \frac{g}{2c_W} 
\bar{\psi}^i_{Q} 
\gamma^\mu 
\Big[ 
X^{QL}_{ij} P_L +X^{QR}_{ij} P_R
-2 s_W^2 Q \delta_{ij} \Big] \psi^j_{Q} Z_\mu, \label{Zcouplings}\\
\mathcal{L}^W &=& \frac{g}{\sqrt{2}}
\bar{\psi}^i_{Q} 
\gamma^\mu
\Big[
V^{QL}_{ij} P_L+V^{QR}_{ij} P_R 
\Big]
\psi^j_{(Q-1)}W^+_\mu + \mathrm{h.c.}~,\label{Wcouplings}
\end{eqnarray}
where $V^Q=0$ for the minimum $Q$ in the model and $P_{L, R} = (1 \mp \gamma_5)/2$ are chirality projectors.
In the SM, $X^{\frac{2}{3}L}_{ij} = - X^{-\frac{1}{3}L}_{ij} = \delta_{ij} $, $V^{\frac{2}{3}L}_{ij} =V^{CKM}_{ij}$ 
and all the other couplings vanish at tree level. 
The couplings to the Goldstone bosons are denoted by
\begin{eqnarray}
\mathcal{L}^{G^0} &=&
\frac{g}{2c_W}\bar{\psi}^i_{Q} 
\Big[
Y^{QL}_{ij} P_L+Y^{QR}_{ij} P_R 
\Big]
\psi^j_{Q}G^0, \label{G0couplings}\\ 
\mathcal{L}^{G^\pm} &=&
\frac{g}{\sqrt{2}}\bar{\psi}^i_{Q} 
\Big[
W^{QL}_{ij} P_L+W^{QR}_{ij} P_R 
\Big]
\psi^j_{(Q-1)}G^+ + \mathrm{h.c.}~. 
\label{Gcouplings}
\end{eqnarray}
Note that we have extracted a factor of $g/2c_W$ and $g/\sqrt{2}$ in
the couplings of the Goldstone bosons to simplify the equations of the
observables. Finally, the trilinear gauge boson and the gauge-Goldstone boson
interactions are those of the SM~\cite{Denner:1991kt}, 
\begin{eqnarray}
 \mathcal{L}^{int}_{g-g,g-G} &=&
 -g \, \bigg\{
   c_W 
    \left[g^{\mu \nu}(k_1-k_2)^{\rho} + g^{\nu \rho}(k_2-k_3)^{\mu} + g^{\rho \mu}(k_3-k_1)^{\nu} \right]
   Z_{\mu}(k_1)W^{+}_{\nu}(k_2)W^{-}_{\rho}(k_3)
  \nonumber \\
  && \phantom{-g} +
  \frac{1-2 s_W^2}{2 c_W} \left(k^{-}-k^{+} \right)^{\mu} Z_{\mu}G^{+}(k^{+})G^{-}(k^{-})
  \nonumber \\
  && \phantom{-g} +
  \left( m_W \frac{s_W^2}{c_W} g^{\mu \nu} Z_{\mu}G^{+}W^{-}_{\nu} + \textrm{h.c.} \right)
 \bigg\},
\label{3bosonvertex}
\end{eqnarray}
where all momenta are taken to flow into the vertices.

\subsection{Result for the $T$ parameter}

The $T$ parameter measures the amount of custodial symmetry breaking
and can be defined in terms of the vacuum polarization
amplitudes of the $SU(2)_L$ $W^{i}_\mu$ gauge bosons as~\cite{Peskin:1991sw}
\begin{equation}
T= \frac{4\pi}{m_Z^2 s_W^2 c_W^2}\big[ \Pi_{+-}(0)-\Pi_{33}(0)\big],
\end{equation}
where $\Pi_{ij}(0)$ denotes the transverse part of the 
vacuum polarization amplitude evaluated at zero momentum,
\begin{equation}
\mathrm{i} g^{\mu\nu}\Pi_{ij}(p^2)+ (p^\mu p^\nu \mbox{ terms}) 
\equiv \int d^4 x\, e^{-i px} \langle J^\mu_i(x) J^\nu_j(0)\rangle.
\end{equation}
The $T$ parameter was computed in~\cite{Lavoura:1992np} for an
arbitrary number of vector-like singlets and doublets. We have
extended their calculation to arbitrary couplings. The final result,
which we reproduce here for completeness, is almost unchanged,
\begin{eqnarray}
\Delta T&=& \frac{N_c}{16\pi s_W^2 c_W^2}
\Big\{
\sum_{\alpha,\beta}
[(|V^{QL}_{\alpha \beta}|^2+|V^{QR}_{\alpha \beta}|^2) 
\theta_+(y_\alpha,y_\beta)
+2 \mathrm{Re}(V^{QL}_{\alpha \beta}V^{QR\ast}_{\alpha \beta})
\theta_-(y_\alpha,y_\beta)
]
\\
&&
\phantom{\frac{N_c}{16\pi s_W^2}}
-\frac{1}{2}\sum_{\alpha,\beta}
[(|X^{QL}_{\alpha \beta}|^2+|X^{QR}_{\alpha \beta}|^2) 
\theta_+(y_\alpha,y_\beta)
+2 \mathrm{Re}(X^{QL}_{\alpha \beta}X^{QR\ast}_{\alpha \beta})
\theta_-(y_\alpha,y_\beta)
]
\Big\}. \nonumber 
\end{eqnarray}
Here again the sum over $Q$ is left implicit; 
we have defined $y\equiv m^2/m_Z^2$, with $m$ the mass of the
corresponding quark, and the functions $\theta_\pm$ read
\begin{eqnarray}
\theta_+(y_1,y_2)&\equiv& y_1+y_2-\frac{2y_1 y_2}{y_1-y_2}\log
  \frac{y_1}{y_2}-2(y_1 \log y_1 + y_2 \log y_2) + \frac{y_1+y_2}{2}
  \Delta, \\
\theta_-(y_1,y_2)&\equiv & 2 \sqrt{y_1 y_2} \left(
\frac{y_1+y_2}{y_1-y_2}\log \frac{y_1}{y_2}-2 
+\log(y_1y_2)-\frac{\Delta}{2}\right),
\end{eqnarray}
where $\Delta$ is a divergent term that comes from dimensional
regularization. We have left it explicit so that it is possible to
check the cancellation of poles in the $T$ parameter. Taking the SM
limit (only $t$ and $b$ quarks) we obtain
\begin{eqnarray}
\Delta T_{SM}&=& 
\frac{N_c}{16 \pi s_W^2 c_W^2} 
\Big\{
\theta_+(y_t,y_b)
-\frac{1}{2}(
\theta_+(y_t,y_t)+\theta_+(y_b,y_b))
\Big\}
\nonumber \\
&=&
\frac{N_c}{16 \pi s_W^2 c_W^2 m_Z^2} 
\Big[ m_t^2+m_b^2- 2 \frac{m_t^2 m_b^2}{m_t^2-m_b^2}
\log\frac{m_t^2}{m_b^2}
\Big] 
\approx 1.19~.
\end{eqnarray}

\subsection{Anomalous  $Zb_L \bar{b}_L$ coupling at one loop}

\begin{figure}[t,b]
\includegraphics[width=0.6\textwidth]{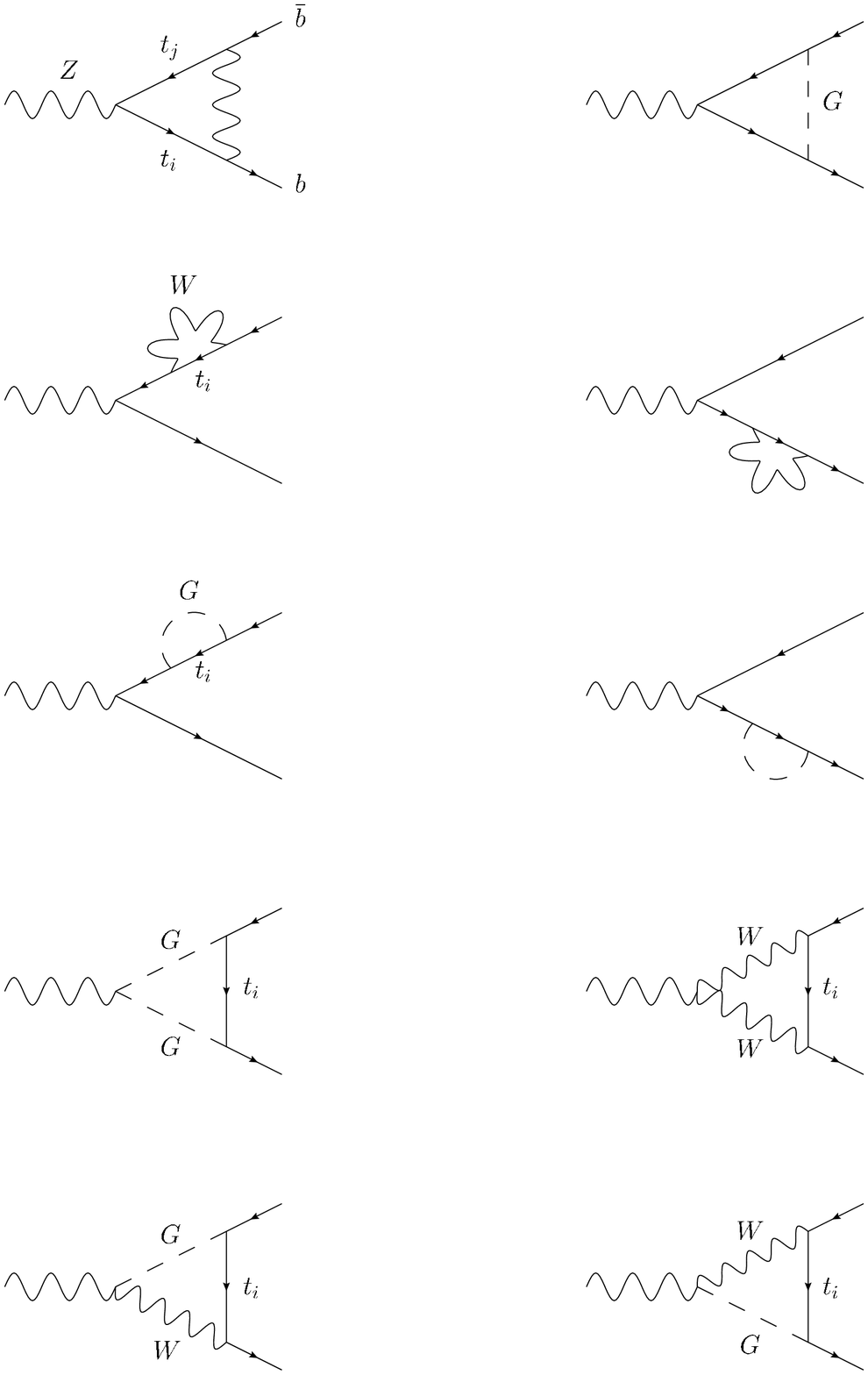}
\caption{Diagrams with heavy quarks contributing  to  the $Z b_L \bar{b}_L$ amplitude.}
\label{fig:Zbbdiags} 
\end{figure}

The amplitude for the decay of a $Z$ boson to massless
left-handed b-quarks,
\begin{equation}
Z(q) \to \bar{b}_L(p_2) b_L(p_1)  ,
\end{equation}
can be written
\begin{equation}
{\cal M}_{Z \to \bar{b}_L b_L} = -\frac{e g_L}{s_W c_W} \bar{b}(p_2) 
\dslash \epsilon(q)
\frac{1- \gamma_5}{2} b(p_1)  ,
\end{equation}
where the $g_L$ coupling can be modified from its value in the SM 
\begin{equation}
g_L = g_L^{{\rm SM}} +  \delta g_L
\end{equation}
due  to new heavy quarks  with $W$ and $Z$ boson interactions described  by the 
Lagrangian of Eqs. (\ref{Zcouplings}-\ref{Gcouplings}).
There are severe constraints on the value of $g_L$ from the LEP experiments. In our 
model, a symmetry protection of this coupling  has been 
implemented to forbid tree level corrections. 
However, modifications can occur via radiative corrections. 
Taking into account one-loop effects we have
\begin{equation}
\delta g_L =  \frac{\alpha}{2 \pi} \left( F^{heavy} -  F^{top}_{SM}\right),
\end{equation}
where we denote with $F^{{\rm heavy}}$ one-loop contributions  from all the heavy 
quarks  in the general BSM model, including the top quark with couplings as in the 
BSM model. The prediction for $g_L^{{\rm SM}}$ includes already contributions from the 
top quark with SM couplings, $F^{{\rm top}}_{{\rm SM}}$; these have been removed 
explicitly in the above equation. 

Summing over all the diagrams in Fig.~\ref{fig:Zbbdiags} and  carrying out  the 
renormalization procedure, we obtain~\footnote{In order to make our result more compact,
 here we already substituted for the trilinear gauge boson and for the gauge-Goldstone boson
 couplings. Therefore, when applying this formula one should be careful in writing all the
 fermion-(gauge/Goldstone) boson couplings appearing in it consistently with the conventions of 
 Eq. (\ref{3bosonvertex}). } 
 
\begin{eqnarray}
F^{{\rm heavy}} & = & - \frac{1}{8 s_W^2} \sum_{i} \Bigg\{
\sum_{j} \Big[
V^{QL}_{jb} V^{QL*}_{ib} \left(2 \tilde{X}^{QR}_{ij}E_1^{ij} +\tilde{X}^{QL}_{ij} E_2^{ij} \right)
\nonumber\\
& & \phantom{wwwwwwy\sum_j} + 
W^{QL}_{jb} W^{QL*}_{ib} \left(\tilde{X}^{QL}_{ij} E_1^{ij} + \tilde{X}^{QR}_{ij} E_3^{ij}\right) 
\Big] \nonumber \\ % end sum over j
& & \phantom{wwwww\sum_j} + \tilde{X}^L_{bb} \big[|V^{QL}_{ib}|^2 (2 E_4^{i}-1) + |W^{QL}_{ib}|^2 E_4^{i} \big] \nonumber \\
& & \phantom{wwwww\sum_j} + (2 s_W^2-1) |W^{QL}_{ib}|^2 E_5^{i} - 2 c_W^2 |V^{QL}_{ib}|^2 E_6^{i} + 4 s_W^2 \textrm{Re}(V^{QL*}_{ib} W^{QL}_{ib}) E_7^{i}
\Bigg\}.
\nonumber\\
\label{AnomalZbb}
\end{eqnarray}
We  have defined  
\begin{equation}
  \tilde{X}^{Q(L,R)}_{ij} = X^{Q(L,R)}_{ij} -2 s_W^2 Q \delta_{ij}, 
\end{equation}
and 
\begin{eqnarray}
E_1^{ij} & = & \sqrt{y_i y_j}\, {\mathcal I}_3 (y_i,y_W,y_j), 
\\
E_2^{ij} & = & \Delta-2+y_i+y_j-2 y_W - {\mathcal I}_2(y_i,y_j) \big(y_i+y_j-2 y_W-3 \big) 
\nonumber \\ 
& & + 2 {\mathcal I}_3 (y_i,y_W,y_j)  \big(y_i-y_W - 1 \big) \big(y_j-y_W-1 \big) + \log(y_i) \left(\frac{2y_i}{y_i-y_W}-y_i\right)   
\nonumber \\
& & + \log(y_j) \left(\frac{2 y_j}{y_j-y_W}-y_j\right) + 2 y_W \log(y_W)\left(1-\frac{ y_i+y_j-2 y_W}{(y_i-y_W)(y_j-y_W)} \right),
\\
E_3^{ij} & = & \frac{1}{2} \Big[ \Delta + 1 + y_i+y_j-2 y_W - {\mathcal I}_2(y_i,y_j) \big(y_i+y_j-2 y_W+1 \big)  \nonumber\\ 
& & + 2 {\mathcal I}_3 (y_i,y_W,y_j)  \big(y_i - y_W \big) \big(y_j - y_W \big) - y_i \log(y_i) - y_j \log(y_j) \nonumber \\ 
& & + 2 y_W \log(y_W) \Big], 
\\
E_4^{i} & = & \frac{1}{2}\left[-\Delta+\frac{1}{2}+\frac{y_i}{y_i-y_W} - \log(y_i)\frac{y_i^2}{(y_i-y_W)^2} + y_W \log(y_W) \frac{2y_i-y_W}{(y_i-y_W)^2}\right],
\\
E_5^{i} & = & \frac{\Delta}{2} - \frac{1}{2} + y_i-y_W - {\mathcal I}_2 (y_W,y_W)  \left(y_i-y_W+\frac{1}{2}\right)  \nonumber\\
& & -  {\mathcal I}_3 (y_W,y_i,y_W) \Big((y_i-y_W)^2+y_i\Big) - y_i \log(y_i) + y_W \log(y_W),
\\
E_6^{i} & = & 3\Delta-4+2 \big(y_i-y_W \big) - {\mathcal I}_2 (y_W,y_W) \big(2 y_i-2 y_W-1 \big)  \nonumber\\
& & - 2 {\mathcal I}_3 (y_W,y_i,y_W) \big( (y_i-y_W)^2+2 y_W \big)  \nonumber\\
& & + 2 \log(y_i) \left(\frac{2 y_i}{y_i-y_W}-y_i \right) + 2 \log(y_W) \left(-\frac{2 y_W}{y_i-y_W}+y_W \right),
\\
\textrm{and} && \nonumber
\\
E_7^{i} & = & \sqrt{y_W y_i}\, {\mathcal I}_3 (y_W,y_i,y_W).
\end{eqnarray}
The finite parts of the 
two-point and  three-point master integrals  can be easily evaluated  numerically 
from their integral representations:
\begin{eqnarray}
{\mathcal I}_2 (y_1,y_2) & = & -\int_0^1 dx\log[x y_1+ (1-x) y_2- x(1-x)], \\
{\mathcal I}_3 (y_1,y_2,y_3) & = & -\int_0^1 dx\frac{1}{x+y_2-y_3}\log\left[\frac{x y_1+(1-x) y_2}{x y_1+(1-x) y_3-x(1-x)}\right].
\end{eqnarray}

The calculation of the anomalous  $Z b_L \bar{b}_L$ coupling required tensor one-loop 
Feynman integrals with up to three propagators with different masses, 
and the external Z boson invariant mass, which we computed using the scalar form factors 
decomposition in~\cite{Davydychev:1991va, Anastasiou:1999bn}  and 
the program AIR~\cite{Anastasiou:2004vj}. We note  that our exact expressions above 
agree with the limits for the $Zb_L \bar{b}_L$ amplitude that have been presented in 
Ref.~\cite{Bamert:1996px}.

The result for $F_{{\rm SM}}^{{\rm top}}$ can be obtained  from the above  results  in the special case 
of $i=j=t$ by substituting appropriately the  SM couplings of the top quark with gauge and  Goldstone bosons. 

In the model presented here the only relevant contribution to the anomalous $Z b_L \bar{b}_L$ 
coupling is given from charge $2/3$ quarks in the loop. Yet, the result we give can be extended
to models with a different quark content. For example, if heavy charge $-1/3$ quarks were
relevant, their contribution to the anomalous $Z b_L \bar{b}_L$ coupling can be obtained from
the first three lines of Eq. (\ref{AnomalZbb}) with the substitutions
\begin{equation}
y \rightarrow y (m_W^2/m_Z^2)\,, \qquad
V^{Q(L,R)}_{ij} \rightarrow \frac{1}{\sqrt{2} c_W} \tilde{X}^{Q(L,R)}_{ij} \qquad
\textrm{and} \qquad
W^{Q(L,R)}_{ij} \rightarrow \frac{1}{\sqrt{2} c_W} Y^{Q(L,R)}_{ij}\, .
\end{equation}

\subsection{Result for $B_q-\bar{B}_q$}

$\Delta B=2$ processes can be conveniently parametrized in terms of
the following dimension 6 effective Lagrangian,
\begin{equation}
\mathcal{L}^{\Delta B=2}
= \sum_{i=1}^5 C_i Q_i^{bq}+\sum_{i=1}^3 \tilde{C}_i \tilde{Q}_i^{bq},
\end{equation}
where we have used standard notation
\begin{eqnarray}
Q_1^{bq}&=&
\bar{q}_L^\alpha \gamma^\mu b_L^\alpha \,
\bar{q}_L^\beta \gamma_\mu b_L^\beta, 
\quad
\tilde{Q}_1^{bq}=
\bar{q}_R^\alpha \gamma^\mu b_R^\alpha \,
\bar{q}_R^\beta \gamma_\mu b_R^\beta, 
\\
Q_2^{bq}&=&
\bar{q}_R^\alpha b_L^\alpha \,
\bar{q}_R^\beta b_L^\beta, 
\quad\quad\quad
\tilde{Q}_2^{bq}=
\bar{q}_L^\alpha b_R^\alpha \,
\bar{q}_L^\beta b_R^\beta, 
\\
Q_3^{bq}&=&
\bar{q}_R^\alpha b_L^\beta \,
\bar{q}_R^\beta b_L^\alpha, 
\quad\quad\quad
\tilde{Q}_3^{bq}=
\bar{q}_L^\alpha b_R^\beta \,
\bar{q}_L^\beta b_R^\alpha, 
\\
Q_4^{bq}&=&
\bar{q}_R^\alpha b_L^\alpha \,
\bar{q}_L^\beta b_R^\beta, \\
Q_5^{bq}&=&
\bar{q}_R^\alpha b_L^\beta \,
\bar{q}_L^\beta b_R^\alpha. 
\end{eqnarray}
We have computed the Wilson coefficients for the different operators
$C_i$, $\tilde{C}_i$, due to the exchange, in box diagrams, of charge
$2/3$ quarks with arbitrary couplings as parametrized in
Eqs.~(\ref{Wcouplings}) and (\ref{Gcouplings}). The procedure requires
Fierz rearragement but is otherwise standard. 
We define the mass ratios 
\begin{equation}
x_i =  \frac{m_i^2}{m_W^2}.
\end{equation}
The final result reads:
\begin{eqnarray}
C_1 &=& \frac{G_F^2 m_W^2}{8\pi^2} 
\sum_{i,j}
\Big[
g_1(x_i,x_j)\left(
\frac{1}{2} 
W^{L\ast}_{is} W^L_{ib} W^{L\ast}_{js} W^L_{jb} 
+ 2 
V^{L\ast}_{i s} V^{L}_{ib} 
V^{L\ast}_{j s} V^{L}_{jb} 
\right) 
\nonumber \\
&& \phantom{G_F^2 m_W^2 WW}
-4 \sqrt{x_i x_j} 
g_0(x_i,x_j) 
W^{L \ast}_{is} V^L_{ib} 
V^{L\ast}_{js}  W^{L}_{jb}
\Big], \\
C_2 &=& \frac{G_F^2 m_W^2}{8\pi^2} 
\sum_{i,j}
2 \sqrt{x_i x_j} g_0(x_i,x_j) 
W^{R\ast}_{is} W^{L}_{ib} W^{R\ast}_{js} W^{L}_{jb},\\
C_3 &=& \frac{G_F^2 m_W^2}{8\pi^2} 
\sum_{i,j}
\Big[
8g_1(x_i, x_j) 
W^{R\ast}_{is} V^L_{ib}
V^{R\ast}_{js} W^{L}_{jb}
-16 \sqrt{x_i x_j} 
g_0(x_i,x_j) 
V^{R\ast}_{is} V^{L}_{ib} V^{R\ast}_{js} V^{L}_{jb}
\Big],\\
C_4 &=& \frac{G_F^2 m_W^2}{8\pi^2} 
\sum_{i,j}
\Big[
4
\sqrt{x_i x_j} 
g_0(x_i,x_j) 
(
W^{R\ast}_{is} W^{L}_{ib} W^{L\ast}_{js} W^R_{jb}
+ 4 
V^{R\ast}_{is} V^L_{ib} V^{L\ast}_{js} V^{R}_{jb}
)
\nonumber
\\
&&  \phantom{G_F^2 m_W^2 WW}
-4g_1(x_i,x_j)
( W^{R\ast}_{is} V^L_{ib} V^{L\ast}_{js} W^{R}_{jb}
+
W^{L\ast}_{is} V^R_{ib} V^{R\ast}_{js} W^{L}_{jb}
)
\Big],
\\
C_5 &=& \frac{G_F^2 m_W^2}{8\pi^2} 
\sum_{i,j}
\Big[
g_1(x_i,x_j) 
\left(
-2
W^{L\ast}_{is} W^L_{ib} W^{R\ast}_{js} W^R_{jb}
-32 V^{L\ast}_{is} V^{L}_{ib} V^{R\ast}_{js} V^{R}_{jb}
\right)
\nonumber \\
&& \phantom{G_F^2 m_W^2 WW}
+ 8  \sqrt{x_i x_j} g_0(x_i,x_j) 
(
W^{L\ast}_{is} V^{L}_{ib} V^{R\ast}_{js} W^R_{jb}
+W^{R\ast}_{is} V^{R}_{ib} V^{L\ast}_{js} W^{L}_{jb}
)
\Big].
\end{eqnarray}
Also $\tilde{C}_i = C_i (L \leftrightarrow R)$.\\
In the above equations the functions $g_0(x, y) $ and $g_1(x, y)$ are
\begin{eqnarray}
 g_0(x, y) & = & -\frac{J_0(x)-J_0(y)}{x-y}, \nonumber \\
 g_1(x, y) & = & -\frac{J_1(x)-J_1(y)}{x-y},
\end{eqnarray}
with
\begin{eqnarray}
 J_0(x) & = & \frac{x \log(x)}{(1-x)^2} + \frac{1}{1-x}, \nonumber \\
 J_1(x) & = & \frac{x^2 \log(x)}{(1-x)^2} + \frac{1}{1-x}.
\end{eqnarray}

\subsection{Results for $b\to s \gamma$}

$B \to X_s \gamma$ is an interesting observable, as it can be very
sensitive to new physics. It probes different combinations
of top couplings than the other observables that we have considered so far,
and in principle it  restricts further the allowed parameter
space in our model. However, because of the same reason,
deviations in $b \to s \gamma$ are not necessarily correlated to those
of $T$ and $\delta g_{b_L}$. This means that while arbitrary points in
parameter space can be constrained by $b \to s \gamma$, small
modifications of other sectors in the model outside the top 
(like for instance details of how the $b$ quark gets a mass) can easily render 
this observable compatible with experimental measurements without modifying the values
of $T$ or $\delta g_{b_L}$.

We use the results of Ref.~\cite{Bobeth:1999ww}, in which the relevant
matching conditions are computed including the NLO QCD corrections in an
arbitrary extension of the SM (however, we use only LO QCD correction in our
estimation). The relevant operators are
\begin{equation}
Q_7=\frac{em_b}{16\pi^2}(\bar{s}_L \sigma^{\mu\nu}b_R) F_{\mu\nu},
\quad
Q_8=\frac{g_s m_b}{16\pi^2}(\bar{s}_L \sigma^{\mu\nu}T^a b_R) G^a_{\mu\nu},
\end{equation}
where $e$ and $g_s$ are the electromagnetic and strong coupling
constants, respectively, $F_{\mu\nu}$ and $G^a_{\mu\nu}$ the
electromagnetic and gluonic field strength tensors and $T^a$ are the
color matrices normalized to $\mathrm{Tr}T^aT^b=\delta_{ab}/2$.
Splitting the corresponding Wilson coefficients into a SM part and a
new physics part, at the matching scale,
\begin{equation}
C_{7,8}=C_{7,8}^\mathrm{SM}+\Delta C_{7,8},
\end{equation}
we can express the constraint on the new physics contribution as:
\begin{eqnarray}
\mathcal{B}(\bar{B}\to X_s \gamma)&=& 
\Big[ 3.15 \pm 0.23
-8.03 \Delta C_7 - 1.92 \Delta C_8
\nonumber \\
&& \phantom{\Big[}
+4.96 (\Delta C_7)^2 + 0.36 (\Delta C_8)^2
+2.33 \Delta C_7 \Delta C_8 \Big] 10^{-4},
\end{eqnarray}
where the SM contribution includes NNLO results~\cite{Freitas:2008vh}. 
This result is to be compared with the
experimental average~\cite{Barberio:2008fa}
\begin{equation}
\mathcal{B}(\bar{B}\to X_s \gamma)_{\mathrm{exp}}
=(3.52 \pm 0.23\pm 0.09) \times 10^{-4}.
\end{equation}

%%%%%%%%%%%%%%%%%%%%%%%%%%%%%%%%%%%%%%%%%%%%%%%%%%%%%%%

\section{Constraints on the fermionic sector and collider
  implications} 
\label{ewpt}
In this section we analyze  the main electroweak and flavor experimental constraints on our 
class of composite Higgs models. 
The fact that we require a very constrained 
but non-negligible contribution from the fermionic sector to the $T$-parameter 
while the other observables must not be disturbed significantly,  
renders the $\chi^2$ quite sensitive to the model parameters. 
This  sensitivity could be  reduced by choosing large values of $f$, since in the
infinite  $f$ limit we recover the SM.  However, this possibility is not attractive because 
large values of $f$ do not help with addressing the usual SM fine-tuning problem. 
For small values, EWSB is less fine-tuned but a larger contribution to $T$ from the
fermionic sector is required to make the model compatible with precision data. 
From now on we will consider as our benchmark scenario $f=500$ GeV. This 
corresponds to the lowest point in Fig.~\ref{STplot} and the model is subject 
to non-trivial constraints (see Fig.~\ref{TgbLplot})
while at the same time its naive fine-tuning measure~\cite{Barbieri:2007bh} is better 
than $\sim 10\%$. 

From Figs.~\ref{STplot} and~\ref{TgbLplot} we see that the region of
parameter space allowed by precision observables is presumably not
only very sensitive to the input parameters but also
small.~\footnote{Whether this constitutes further fine-tuning is a
  debatable issue. The fermionic sector will also contribute, in a
  full composite Higgs model, to the Higgs potential and a
  sensible measure of fine-tuning might be the overlap between the
  regions with good EWSB and good compatibility with precision
  data. Ref.~\cite{Panico:2008bx} presented a 5D composite Higgs model
  in which the two regions overlap nicely.} 
We have found  that in such a situation, 
computing the electroweak observables exactly or estimating them 
in the large Yukawa approximation can lead to important phenomenological differences. 
In order to estimate this effect and to ensure probing
all relevant regions in parameter space, we have performed several
scans, based on adaptive Monte-Carlo methods~\cite{Hahn:2004fe}, 
which we require to search for phase-space regions with a  small value of $\chi^2$. 
We have implemented two types of scans: in the first type the $\chi^2$ is 
obtained using the complete one-loop calculation of the electroweak observables 
and in the  second  type the $\chi^2$ is obtained from the calculation of
$\delta g_{b_L}$ in the large Yukawa approximation.

In our scans we have restricted $|\mu_{ij}| \lesssim 4 \pi$. We have
also checked that the mass
parameters $m_{L,R}^i$  and $m_\Psi^i$ are typically below $\Lambda$
in the case of two
multiplets (but either $m_\Psi$ or $m_L$ are close or above $\Lambda$ in most of the parameter
space for one multiplet)
and the masses of the fermions that \textit{affect} the relevant
observables are also typically below
the cut-off of our effective theory.

\subsection{One multiplet}

We now discuss the constraints imposed by electroweak and
flavor precision data in the case that there is only one fermionic 
$(5)$ of $SO(5)$ below the cut-off of our effective theory. This case
has been previously considered in the literature, in the same or
similar models
(see~\cite{Lodone:2008yy,Gillioz:2008hs,Pomarol:2008bh}). 
The new  result of our paper is an exact treatment at one loop of all 
the important precision observables; we also elaborate further on the 
implications of this minimal model for collider phenomenology 
(Ref.~\cite{Pomarol:2008bh} also emphasized the collider implications of a composite top). 
We view  the case of one multiplet in this section as  a preparation 
for the more interesting  case of two multiplets  in the following section, 
where we  will discuss  in detail the role  of  a non-minimal  sector  of 
fermionic composites below the cut-off.

\begin{figure}[h,t,b]
\psfrag{dgb}[t][t][.75]{$10^3 \delta g_{b_L}$}
\psfrag{sL}[t][t][.75]{$s_L$}
\includegraphics[width=.45\textwidth]{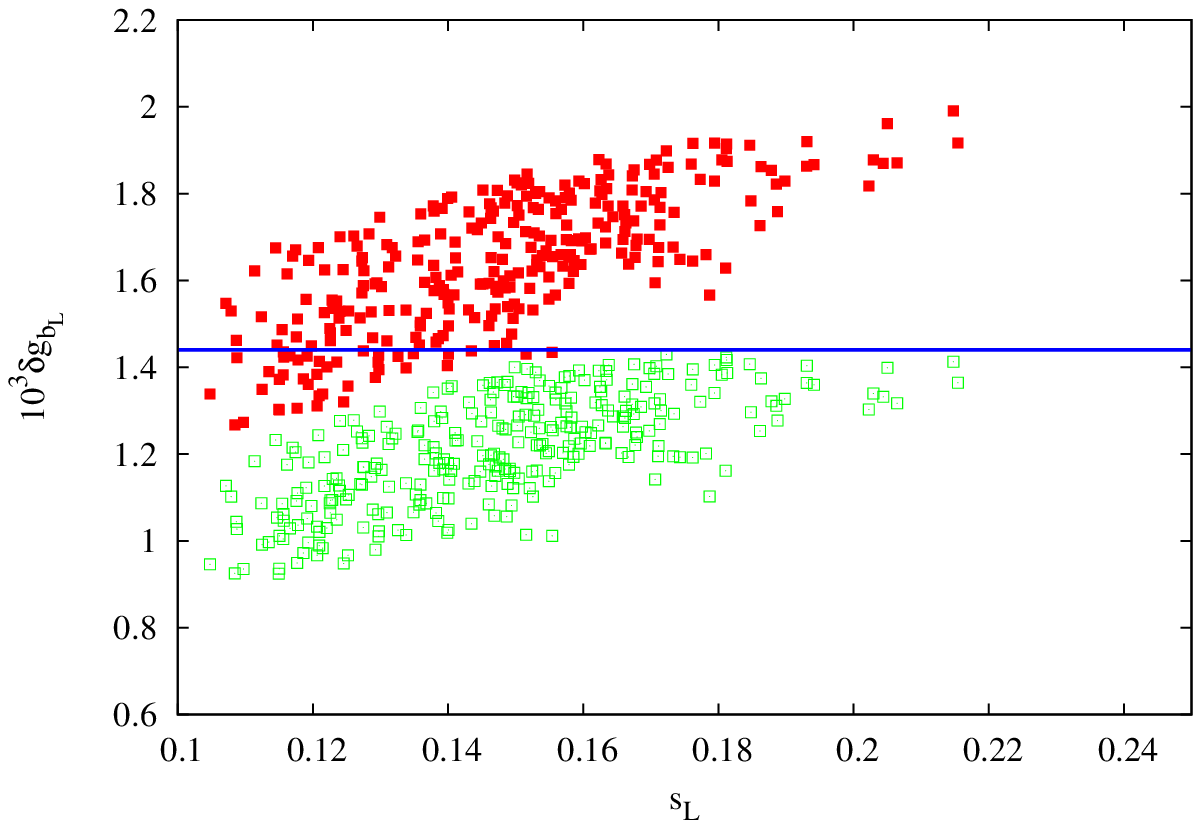}
\hspace{.08\textwidth}
\includegraphics[width=.45\textwidth]{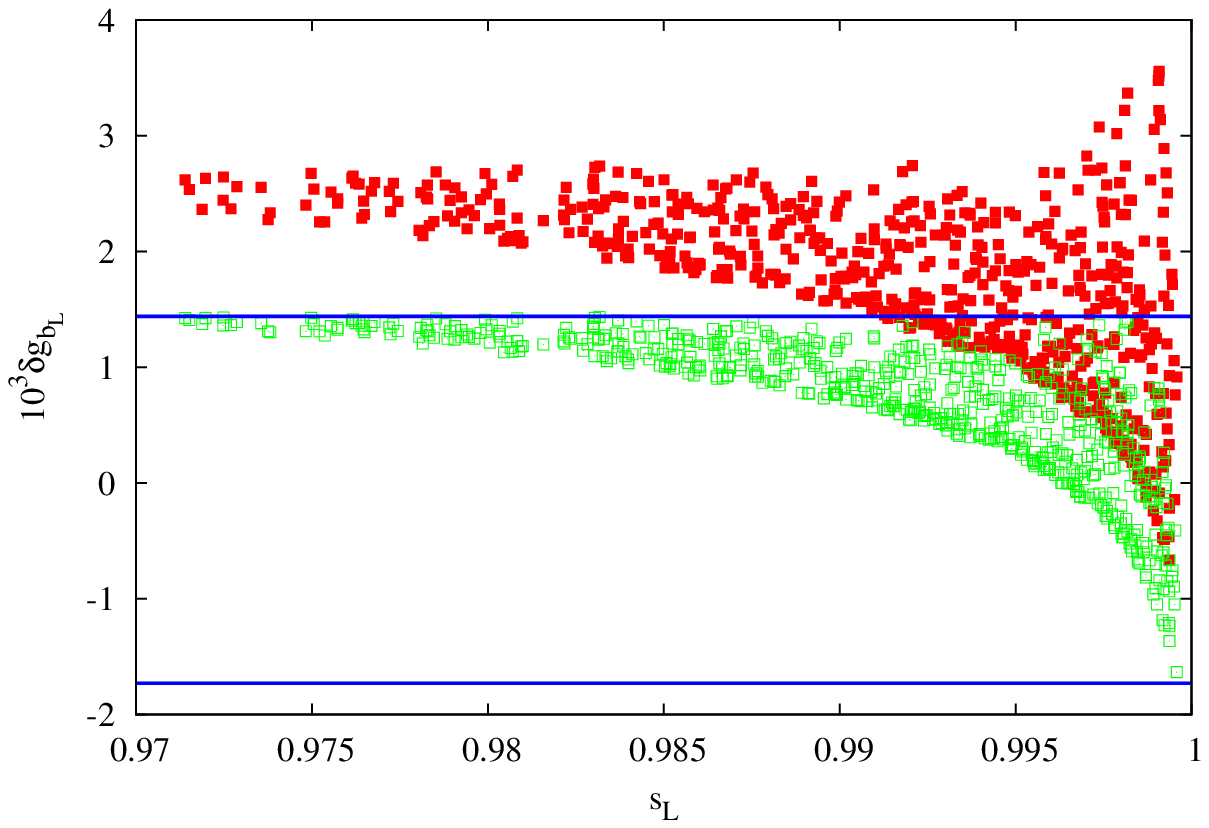}
\caption{ 
 Contribution to $\delta g_{b_L}$ as a function of $s_L$. The other
 input parameters are left free (keeping the total $\chi^2$ using the 
 estimation for $\delta g_{b_L}$ within $99\%$ C.L.).
 The full red (empty green) squares correspond to the full contribution (large Yukawa
 estimation) of $\delta g_{b_L}$. The horizontal lines correspond to
 the maximum allowed contribution to $\delta g_{b_L}$ (assuming an
 optimal contribution to $T$).
}
\label{TvsY} 
\end{figure}

We first show that it is  often  important to use an exact one-loop calculation 
for $\delta g_{b_L}$. We start by performing a  scan of  the parameter space 
using the large Yukawa approximation.   As expected for the case of one multiplet, 
the allowed parameter space is  quite small. 
For $f=500$ GeV, using the large Yukawa coupling approximation, we find two
generic regions in parameter space compatible with electroweak
precision data at the $99\%$ C.L.: one for $0.1 \lesssim s_L \lesssim
0.2$  and one for $s_L\sim 1$. These  regions  are  shown with empty (green) squares 
in the plots of  Fig.~\ref{TvsY}. 
The region  $0.1 \lesssim s_L \lesssim 0.2$ (left plot in Fig.~\ref{TvsY}) is, however,  a misleading artifact of the large Yukawa coupling approximation. When we repeat the  computation of $\delta g_{b_L}$ exactly at one loop, corresponding  to the 
full (red) squares in Fig.~\ref{TvsY}, this region is excluded at 
the $99\%$ C.L. (the region survives only at the $99.9\%$ C.L.). 
On the  other hand, the large $s_L$ region (right plot in Fig.~\ref{TvsY}) 
survives when the full one-loop calculation is used, but  it turns out to be 
significantly smaller than what the analysis using the large Yukawa coupling
approximation indicates. 

\begin{figure}[h,t,b]
\psfrag{dgb}[t][t][.75]{$10^3 \delta g_{b_L}$}
\psfrag{sL}[t][t][.75]{$s_L$}
\includegraphics[width=.65\textwidth]{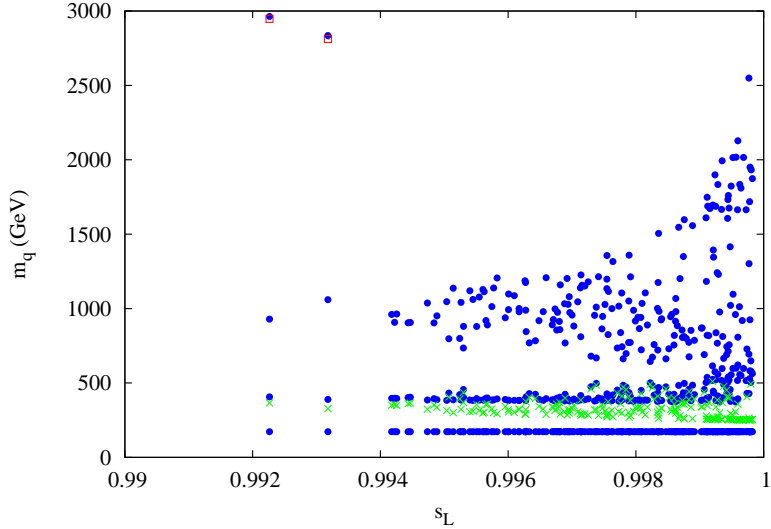}
\caption{ 
Spectrum of light (below $\Lambda$) 
fermionic states (including the top quark) for the region
of parameter space compatible with EWPT. 
The (green) crosses, (blue) dots and (red) empty squares correspond to charge
$5/3$, $2/3$ and $-1/3$ quarks, respectively. The latter, together with one charge $2/3$ quark, are typically
above $\Lambda$.
}
\label{mqbelow3TeV} 
\end{figure}

In Fig.~\ref{mqbelow3TeV}, we show the fermionic spectrum 
below the cut-off of the theory $\Lambda \sim 3$~TeV, in  
the region of parameter  space which is compatible with electroweak 
precision data.
Above the top, there is always a very light ($m_{\frac{5}{3}} \lesssim
500$ GeV) charge $5/3$ quark, then a charge $2/3$ quark very 
close in mass to it ($0\leq m_{\frac{2}{3}}^{(1)}-m_{\frac{5}{3}} \lesssim
100$ GeV for $m_{\frac{5}{3}} \gtrsim 300$ GeV) and finally a heavier
charge $2/3$ quark ($800\mbox{ GeV} 
\lesssim m_{\frac{2}{3}}^{(2)} \lesssim 2$ TeV). The other two quarks, with electric 
charges $2/3$ and $-1/3$ respectively, are quite degenerate and typically heavier
than $\Lambda$. 

\begin{figure}[h,t,b]
\psfrag{Xtt}[t][t][.75]{$X_{tt}$}
\psfrag{VtbR}[t][t][.75]{$|V^R_{tb}|$}
\psfrag{VtbL}[t][t][.75]{$|V^L_{tb}|$}
\psfrag{sL}[t][t][.75]{$s_L$}
\includegraphics[width=.45\textwidth]{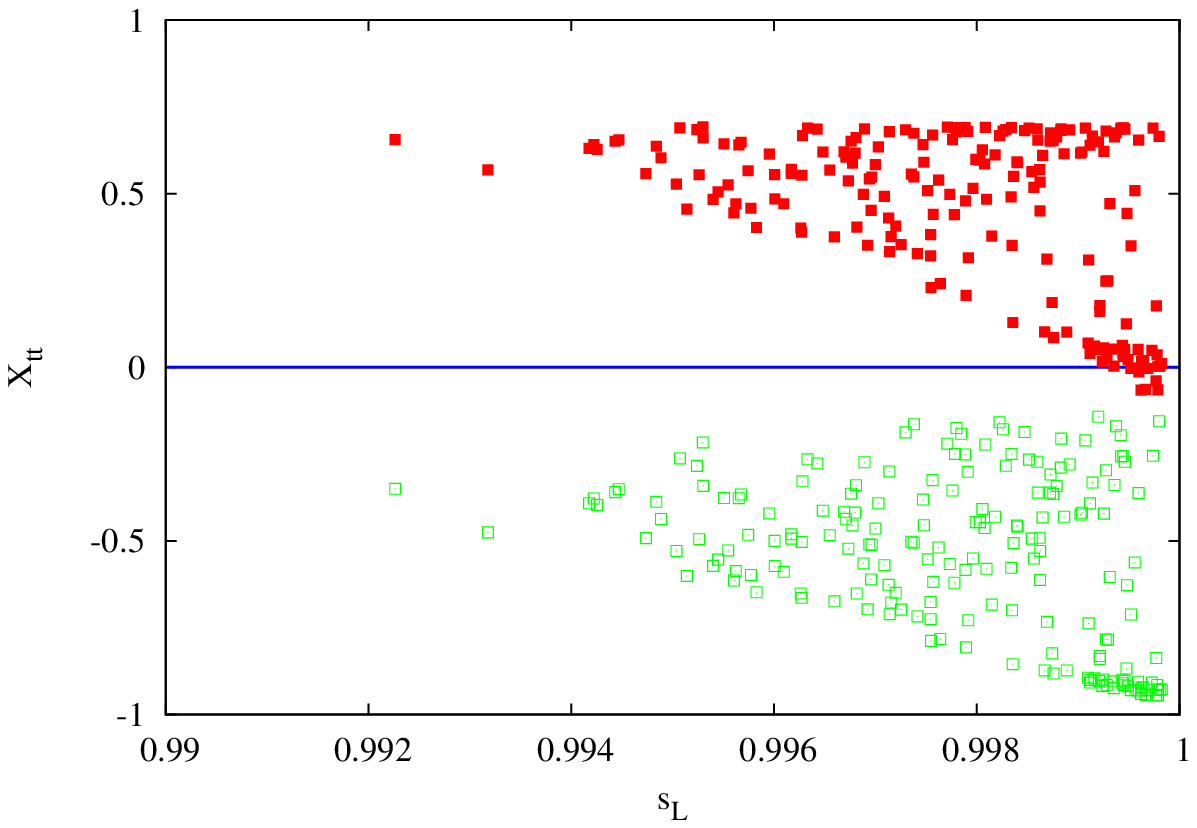}
\hspace{.08\textwidth}
\includegraphics[width=.45\textwidth]{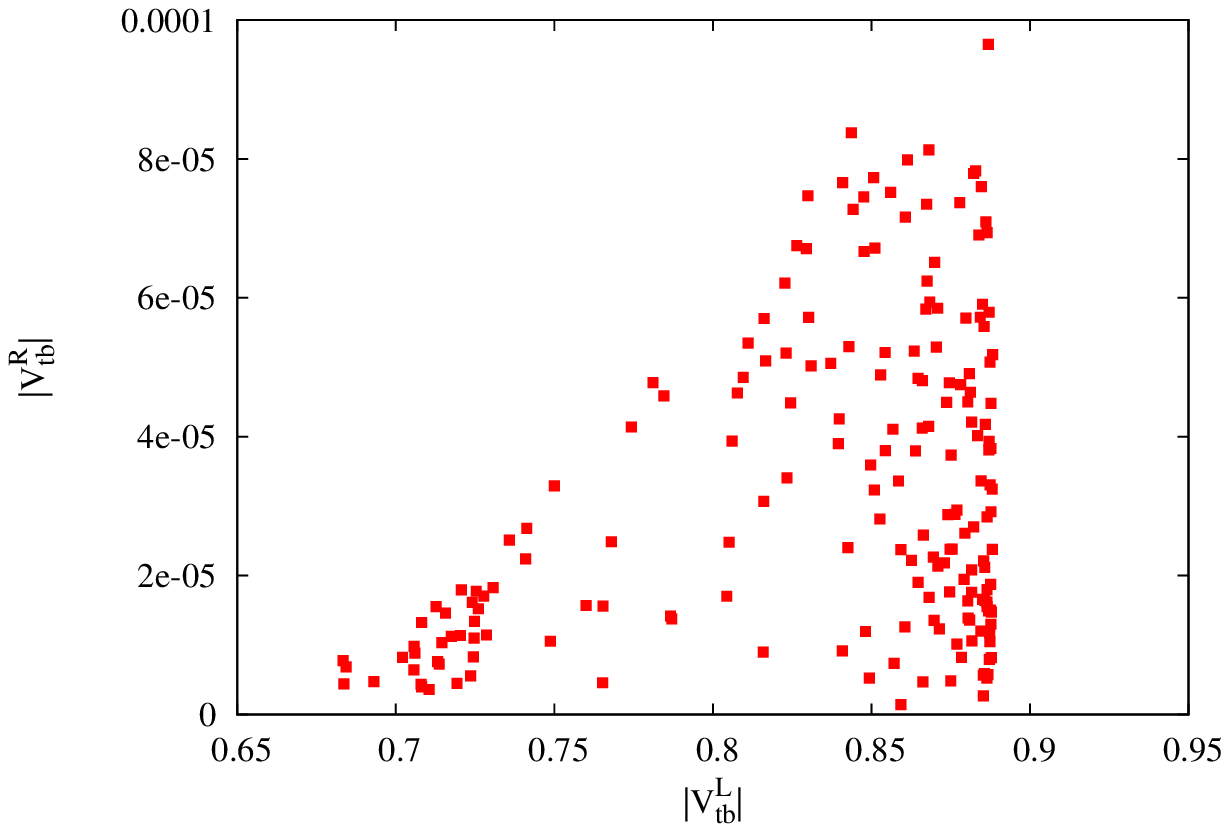}
\caption{ 
$X_{tt}^L$ (empty green squares) and 
$X_{tt}^R$ (full red squares) as a function of $s_L$ (left panel) 
and $V^R_{tb}$ vs $V^L_{tb}$ (right panel). In both cases all the points
are compatible with precision data at the $99\%$ C.L. 
The SM values for these couplings are $X^L_{tt}=V^L_{tb}=1$ and
$X^R_{tt}=V^R_{tb}=0$.}
\label{XttVtb} 
\end{figure}

All these fields mix very strongly with the top and
among themselves to provide the required positive contribution to the
$T$ parameter without violating the bounds on $\delta g_{b_L}$.
This gives rise to large corrections to the top gauge
couplings, $V^{L}_{tb}$ and $X^{L,R}_{tt}$ (the correction to
$V^R_{tb}$ depends on the details of how the bottom quark is embedded in
the theory but they are expected to be suppressed by Yukawas of the
order of the bottom Yukawa). 
We show in Fig.~\ref{XttVtb} the values of these couplings in the
allowed region of our model.
It is interesting that while these couplings receive 
large corrections, the indirect constraints on them from
radiative corrections to electroweak and flavor precision data are
still satisfied thanks to the extra contribution of the new quarks in 
the theory. The charged current top-bottom coupling is only now 
starting to be constrained by single top
production measurements at the Tevatron~\cite{Abazov:2008sz} (see also
\cite{JAAS:talk})
\begin{equation}
|V^L_{tb}| \gtrsim 0.66,
\end{equation}
where we have assumed (as it happens in our model) that the only sizable
correction to the $W t b$ vertex is in $V^L_{tb}$. From
Fig.~\ref{XttVtb} we see that this precision is not yet sufficient in order 
to constrain our model.
The situation will improve at the LHC, where this coupling 
can be determined  with an  accuracy
of about  $\sim 10\%$~\cite{AguilarSaavedra:2008gt} and therefore the model could
be probed not only through direct production of the new quarks but
also in single top production. The measurement  of  the $Z t \bar{t}$ coupling is much
more difficult. It is currently unconstrained and although measurements of 
$Z t \bar{t}$ production at the LHC can be in principle used to measure it, we 
have checked that the achievable precision is likely insufficient to 
constrain our model~\cite{Baur:2005wi,Lazopoulos:2008de}.  

Let us now turn to the collider implications of the new fermionic
sector. Pair production of new vector-like quarks with quantum
numbers similar to ours has been discussed in detail in the
literature~\cite{Carena:2006bn,Contino:2006qr,
Panico:2008bx,Pomarol:2008bh,Cacciapaglia:2006gp,Contino:2008hi,Brooijmans:2008se}. In particular, it
was shown in~\cite{Contino:2008hi,Brooijmans:2008se} that pair
production of charge $5/3$ quarks as light as the ones in our model
can be discovered in the very early runs of the LHC. The lightest of
the charge $2/3$ quarks has sizable couplings to the top, the bottom
and the charge $5/3$ quark. 
Thus, pair production of this quark will result in many $ZZt\bar{t}$ and
$W^+ W^- b \bar{b}$, which should allow for a relatively easy
search. However, the mass difference with the charge $5/3$ quark 
will typically be too small to allow for a significant 
fraction of cascade decays involving both quarks. 

Similar properties are shared by the second charge $2/3$ quark
(order one coupling to all the lighter modes) with the notable distinction 
that now the mass difference with the charge $5/3$ quark is large enough to
allow for a large branching ratio. Thus, we have a sizable set of
events with the spectacular signature 
\begin{equation}
pp \to q^{(2)}_{2/3} \bar{q}^{(2)}_{2/3} 
\to q_{5/3} \bar{q}_{5/3}W^- W^+  
\to W^+ W^+ W^+ W^- W^- W^- b \bar{b}.
\end{equation}
Note however that the larger mass of this quark will significantly
reduce the pair production cross section. A detailed analysis of the
signal and background is necessary to decide the reach of the LHC on
this channel. 

The large couplings of all these fields to the top and
bottom indicates that they can be further tested through single 
production (for a discussion of single production of
vector-like quarks with these quantum numbers mixing with valence
quarks at the Tevatron see~\cite{Atre:2008iu}).

\subsection{Two multiplets}

We have seen in the previous subsection that the allowed region of
parameter space compatible with experimental data is quite small if
only one set of fermionic composites is below the cut-off of our
effective theory. 
Due to the explicit breaking of the $SO(5)$ symmetry 
induced by the mixing with the fundamental fields of the SM top sector,
these composite fermions also contribute to the Higgs potential. 
Naturalness suggests that, ideally, they should provide the
leading contribution to the effective potential, 
cutting-off the Higgs mass not at $\Lambda$ but
rather at the mass of these fermionic resonances. However,
given the severe constraints that precision data impose on the model, 
it would be rather coincidental if the permitted small 
region in parameter space  generated also a viable pattern of
electroweak symmetry breaking with the  correct vacuum expectation value. 
This problem has already been observed in 5D models of composite Higgs, 
in which a non trivial fermionic spectrum was required in order to obtain 
a successful realization of EWSB~\cite{Contino:2006qr,Panico:2008bx}. 

As  a first step in preparation of realistic composite Higgs models which
fully incorporate a satisfactory pattern of EWSB, we find it useful 
to study the effect on electroweak and flavor precision data of a second 
set of light composite fermions. Note that here we do not mean to simply include
the effect of \textit{more Kaluza-Klein modes} (in an analogous 5D picture) 
or heavier resonances in a purely 4D picture.  We are rather considering the 
possibility that the spectrum of fermionic resonances at the scale of 
our strongly coupled  theory is richer than what we have considered 
so far.~\footnote{This is precisely what happens in realistic
5D composite Higgs models. For a 4D interpretation, along the lines of
the models discussed in this paper, see~\cite{Contino:2004vy} and \cite{Contino:2006nn}.}

We go back to our original mass term, Eq.~(\ref{eq:mass-matrixF}), in which now
$m_L$ and $m_R$ are two-dimensional vectors and $m_\Psi$ and $\mu$ are
$2 \times 2$ matrices (the former diagonal, the latter hermitian). The
generalization of $s_{L,R}$ to measure the degree of top compositeness
in the presence of more than one composite is
\begin{equation}
s_L^\textrm{eff}\equiv \sqrt{1-(U^{q(0)}_{11})^2},
\quad
s_R^\textrm{eff}\equiv \sqrt{1-(U^{t(0)}_{11})^2},
\end{equation}
where $U^{q(0)}$
 ($U^{t(0)}$) is the $3\times 3$ unitary matrix
that mixes $q_L$, $Q^1_L$ and $Q^2_L$
($t_R$, $T^1_R$ and $T^2_R$) in order to make the
mass matrix diagonal before EWSB. This definition generalizes to an
arbitrary number of extra composites.

The result of our scans in this case shows that the region of
parameter space compatible with electroweak and flavor precision data
expands dramatically. This was to be expected, due to the increase in
the number of degrees of freedom. Nevertheless, it should be
emphasized that the constraints from the pattern of $SO(5)/SO(4)$
global symmetry breaking are still imposed on our extended sector. Even more
interesting is the fact that the allowed parameter space not
only is it larger,  but we also find a plethora of patterns of 
phenomena, some of which we discuss below. 

The most important phenomenological feature that is allowed by
experimental data when two sets
of composite fermions are below $\Lambda$ is a much 
\textbf{richer spectrum}. The number and couplings of light quarks are 
no longer fixed. We can have from just one single charge $2/3$ quark to a full,
almost degenerate, set of quarks arranged in a $(4)$ of $SO(4)$. They
can be very strongly coupled to the top and bottom or almost not
coupled at all. This has a number of interesting consequences, that
can be subjects for exploration at colliders:
\begin{description}
\item{-\textbf{Complex collider signatures}.} The presence of a
  relatively large, sometimes quite degenerate, set of new particles will 
  require sophisticated analysis to disentangle the contribution of different
  modes. Also, the possibility of long decay chains ending in a large
  number of leptons, jets and missing energy can make it more
  difficult to distinguish composite Higgs models from other new
  physics models in the early data.
\item{-\textbf{Importance of single heavy quark production}.}
Single production, which tests directly the couplings of the new
quarks to the top and bottom, becomes an important tool,
complementary to pair production and to indirect tests through
electroweak precision observables, to fully reconstruct the fermionic
sector of the model.
\item{-\textbf{Strongly composite $t_R$}.} A strongly composite $t_R$
  (as opposed to a strongly composite $t_L$ as in the case of one
  multiplet) is now a quite common
  occurrence. This is a welcome feature, as there are some UV
  completions for which extra sources of flavor violation can 
  be enhanced for a strongly composite $t_L$. 
\end{description}
 
We now discuss in a bit more detail some indicative spectrum patterns 
of new quarks that we have found in our scans.
One interesting possibility is the presence of a rather light ($m_{(4)}\sim
500$ GeV) full, almost degenerate, $(4)$ of $SO(4)$ as the only set of
particles below 1 TeV. This set of new quarks does not 
contribute significantly to the $T$ parameter or $\delta g_{b_L}$ 
(as it barely breaks the custodial symmetry), a role that is mainly played by 
heavier modes which are more difficult to produce at the LHC. The light quarks 
(two charge $2/3$, one $5/3$ and one $-1/3$) all decay mainly to the top and
therefore a large number of $VV t \bar{t}$ events should be produced
at the LHC (Refs.~\cite{Contino:2008hi,Brooijmans:2008se} showed that the charge $5/3$ quarks
should be easily discovered with very early data at the LHC). Once
their masses are known and some information on their couplings from
single production is gathered, a reanalysis of electroweak precision
data should give a clear picture of the fermionic content of the model.

The possibility we have just mentioned, a light full $(4)$ representation of
$SO(4)$, would clearly point to the underlying symmetry structure of
the theory. There are however other regions of parameter space in
which this structure is not so obvious. One example is the case in
which the only light mode, easily accessible at the LHC, is a charge
$2/3$ quark. Of course, we know that in the context of composite Higgs
models with little fine-tuning, a single charge $2/3$ singlet is not
compatible with electroweak data. 
However, without additional information, it will not be easy to 
disentangle the contribution of Higgs compositeness and the new quark 
to electroweak precision data.
Our study of electroweak constraints shows however that the
heavier modes must significantly contribute to the $T$-parameter  
and $\delta g_{b_L}$ and therefore should have large couplings to the top 
and/or bottom that would make them accessible through single production. 

A variety of viable spectra  allows for the possibiliy in which we 
have several non-degenerate modes below 1 TeV. For instance, there are 
regions in which the lightest new quark is a charge $5/3$ one, then a heavier
charge $2/3$ and then a heavier (but still relatively light, $m_{-1/3}
\sim 800$ GeV) charge $-1/3$. The mass difference is commonly large
enough to allow for cascade decays that end in the top and up to three
gauge bosons. This means we can easily have up to eight gauge bosons
and two b's in the final state,
\begin{eqnarray}
pp &\to& q_{-\frac{1}{3}}\bar{q}_{-\frac{1}{3}} 
\to W^- q_{\frac{2}{3}} W^+ \bar{q}_{\frac{2}{3}} 
\to W^- W^- q_{\frac{5}{3}}W^+W^+ \bar{q}_{\frac{5}{3}}
\nonumber \\ 
&\to& W^- W^- W^+ t W^+W^+ W^- \bar{t}
\to W^- W^- W^+ W^+ b W^+W^+ W^- W^- \bar{b}.
\end{eqnarray}
A fraction of the time $q_{\frac{2}{3}}$ will decay into $Z t$ so that
the final state can also be $6W+Z+b\bar{b}$ or  
$4W+2Z+b\bar{b}$. Thus, we have processes with a relatively large
production cross section (for quarks which are not too heavy) 
and a very complex final state with many jets, leptons and missing energy. Although a
detailed analysis is required to assess our capability of 
understanding these complex processes, it is likely that they will be
easy to discover but very difficult to fully reconstruct and the detailed 
information on the quark masses may not be extracted.

We note that the same features as we find here, 
a rich spectrum of light modes, some of which do not
contribute very strongly to electroweak observables, and a number of
heavier modes (but still lighter than the bosonic resonances) which
contribute to render electroweak observables compatible with
experimental data has been recently observed in composite Higgs models
in five dimensions. In fact, this kind of spectrum proved crucial to
make the model in~\cite{Panico:2008bx} \textit{simultaneously}
compatible with electroweak precision tests and with a realistic
pattern of electroweak symmetry breaking.

%%%%%%%%%%%%%%%%%%%%%%%%%%%%%%%%%%%%%%%%%%%%%%%%%%%%%%%

\section{Conclusions}
\label{sec:conclusions}
The realization of electroweak symmetry breaking is a long-standing
mystery that will be soon tested in detail at the LHC. Composite Higgs
models, in which the Higgs boson arises as the (composite)
pseudo-Goldstone boson of a spontaneously broken global symmetry in a
strongly coupled gauge theory, is an appealing candidate. It naturally
protects the electroweak scale from short distance physics and can
even explain the suppression of the electroweak scale with respect to
the scale of new physics. Compatibility with current experimental data
seem to point to a relatively large scale of new bosonic resonances
$\Lambda~\gtrsim~3$~TeV and to a custodially preserving symmetry
breaking pattern to protect the $T$ parameter. Similarly, a left-right
symmetry within the custodial symmetry naturally protects the
$Z b_L \bar{b}_L$ coupling from large corrections. A minimal symmetry
breaking pattern that contains the Higgs as a pseudo-Goldstone boson,
is custodially symmetric and can protect the $Z b_L \bar{b}_L$ coupling is
a global $SO(5)$ symmetry spontaneously broken to $SO(4)$ and fermions
in the fundamental representation of 
$SO(5)$~\cite{Agashe:2004rs,Agashe:2005dk,Agashe:2006at}.

The low energy implications of this set-up can be simply analyzed with
the aid of an effective description based on an $SO(5)/SO(4)$
non-linear sigma model. If the Higgs is quite composite, \textit{i.e.}
if the scale of the global symmetry breaking is not far from the
electroweak scale, as one would expect in a natural (non fine-tuned)
model, its couplings to the Standard Model fields are significantly reduced.
This results in a sensitivity to the cut-off of the theory whose
effect can be taken into account by defining an effective (heavier)
Higgs mass. This effective Higgs mass is the one that should be
included in the calculation of electroweak precision observables,
giving a positive contribution to the $S$ parameter and a negative
contribution to the $T$ parameter. These contributions simply come from
the fact that the Higgs is composite and are therefore quite generic. UV
physics, being custodially symmetric, is not expected to contribute to
the $T$ parameter but it can give a tree level contribution to the $S$
parameter that, together with the contributions to $S$ and $T$ from
Higgs compositeness, put the model in gross contradiction with current
experimental tests. 

The large mass of the top quark, however, makes it natural to assume
that it is partially composite. In that case it will strongly couple 
to the fermionic resonances of the strongly coupled theory which, if
lighter than the cut-off of the theory, can have an important impact
on electroweak observables (due to their mixing with the top, which
breaks explicitly the custodial symmetry). We have considered the
presence of one or more sets of fermionic resonances, spanning full
fundamental representations of $SO(5)$. The $SO(5)$ symmetry is
explicitly broken by the SM quarks, $q_L$ and $t_R$ (the former also
breaks the custodial $SO(4)$ symmetry), which couple linearly to the
strongly coupled theory. This coupling makes $q_L$ and $t_R$ partly
composite and, through their composite components, couple to the Higgs
and get a mass.
We have computed the exact contribution of this new sector to the relevant
electroweak precision observables, $T$ and $\delta g_{b_L}$, and also
to some of the flavor observables that are strongly correlated to this
new fermionic sector, $B_q-\bar{B}_q$ mixing and $b\to s \gamma$. 
We have performed an exact one-loop calculation which does not rely 
on any approximation and presented the results in a general enough 
way that can be easily extended to models beyond the one we have considered. Our
exact calculation, that goes beyond the large Yukawa approximation
which is commonly used to estimate the contribution to the anomalous $Z b_L \bar{b}_L$ 
coupling, can have an important impact if the model is
strongly constrained, as happens in our case in some regions of
parameter space, or if one wants to do precision analysis of new
physics, as it will become necessary if new quarks are discovered 
at the LHC.

Armed with these detailed calculations of the most relevant
electroweak observables, we have used adaptive
Monte-Carlo methods to scan the 
parameter space in search of the regions allowed by experimental
data. The result depends dramatically on whether there exist 
just one or more fermionic multiplets below the cut-off of our theory. 

The case of only one multiplet below $\Lambda$ is quite
constrained. When all the experimental constraints are taken into
account, only a small region with a very composite LH top survives. In
this region the spectrum of new quarks and their couplings are almost
univocally determined by electroweak and flavor precision data. Above
the top, there are typically two quite light ($\lesssim 500$ GeV) 
quarks of charge $5/3$ and $2/3$, respectively (the former typically
slightly lighter) followed by another charge $2/3$ quark with mass
$800$ GeV $\lesssim m_{\frac{2}{3}}^{(2)} \lesssim 2$ TeV. The lighter two quarks should
be easily produced at the LHC and have large enough couplings to the
top to make single production an interesting channel to
study. Furthermore, they typically induce large enough corrections to
the $V_{tb}$ coupling to be detectable at the LHC. The strong constraints on
this possibility and the fact that we have simply
assumed a realistic pattern of electroweak symmetry breaking - which
is fully calculable and therefore imposes further constraints in a UV
completion of the model - have motivated us to consider the possibility
that more than one set of fermionic resonances contribute to
electroweak observables. This motivation is reinforced by the
experience with 5D UV completions of the model, in which electroweak
symmetry breaking imposes non-trivial constraints on the low energy
spectrum, usually requiring a more complex spectrum of light fermionic
resonances than the one we have found in the case of one multiplet.

The situation dramatically changes when two fermionic multiplets
are allowed to contribute to electroweak precision observables. The
spectrum of light resonances is no longer constrained; we have found
from one single quark of charge $2/3$ to four quarks in a full
degenerate multiplet $(4)$ of $SO(4)$ below 1 TeV. Their couplings to
the top can also vastly change, which makes the study of single
production even more interesting. When pair-produced, these new quarks
can produce long decay chains that contain a large number (up to
eight) of SM electroweak vector bosons and a $b\bar{b}$ pair. Thus,
final states with many jets, leptons and missing energy would be a
common signature in these models. 

We conclude that composite Higgs models with no bosonic 
resonances (apart from the Higgs itself) below the 
cut-off of the low energy effective theory
can be fully compatible with current experimental constraints provided
a quite rich spectrum of light fermionic resonances is present in the
model. These new fermionic resonances should be easily produced at the
LHC and would most likely be the first signal of new physics beyond
the SM. Establishing the symmetry pattern from the fermionic spectrum
can however prove more difficult and will require a detailed analysis
of the full experimental information, including pair production,
single production and a detailed analysis of electroweak and flavor
observables. This is really important as other signatures that would
definitely pin down the model as a composite Higgs model, like the
measurement of the Higgs couplings, longitudinal gauge boson
scattering~\cite{Falkowski:2007iv} and the production of new bosonic resonances of the strongly
coupled theory can take much longer at the LHC~\cite{Agashe:2006hk}.

%%%%%%%%%%%%%%%%%%%%%%%%%%%%%%%%%%%%%%%%%%%%%%%%%%%%%%%

\section*{Acknowledgements}
We would like to thank J.A. Aguilar-Saavedra, R. Contino, M. Gillioz, U. Haisch , A. Pomarol, E. Pont\'on and Z. Kunszt for useful discussions.
This work was supported by the Swiss National Science Foundation under
contract 200021-117873.
%

%%%% bibliography %%%%

%%%%%%%%%%%%%%%%%%%%%%%%%%%%%%%%%%%%%%%%%%%%%%%%%%%%%%%


\begin{thebibliography}{99}

%\cite{Kaplan:1983fs}
\bibitem{Kaplan:1983fs}
  D.~B.~Kaplan and H.~Georgi,
  %``SU(2) X U(1) Breaking By Vacuum Misalignment,''
  Phys.\ Lett.\  B {\bf 136} (1984) 183;
  %%CITATION = PHLTA,B136,183;%%
%\cite{Kaplan:1983sm}
%\bibitem{Kaplan:1983sm}
  D.~B.~Kaplan, H.~Georgi and S.~Dimopoulos,
  %``Composite Higgs Scalars,''
  Phys.\ Lett.\  B {\bf 136} (1984) 187.
  %%CITATION = PHLTA,B136,187;%%

%\cite{Contino:2003ve}
\bibitem{Contino:2003ve}
  R.~Contino, Y.~Nomura and A.~Pomarol,
  %``Higgs as a holographic pseudo-Goldstone boson,''
  Nucl.\ Phys.\  B {\bf 671} (2003) 148
  [arXiv:hep-ph/0306259].
  %%CITATION = NUPHA,B671,148;%%

%\cite{Manton:1979kb}
\bibitem{Manton:1979kb}
  N.~S.~Manton,
  %``A New Six-Dimensional Approach To The Weinberg-Salam Model,''
  Nucl.\ Phys.\  B {\bf 158} (1979) 141;
  %%CITATION = NUPHA,B158,141;%%
%\cite{Hosotani:1983xw}
%\bibitem{Hosotani:1983xw}
  Y.~Hosotani,
  %``Dynamical Mass Generation By Compact Extra Dimensions,''
  Phys.\ Lett.\  B {\bf 126} (1983) 309;
  %%CITATION = PHLTA,B126,309;%%
%\cite{Hosotani:1983vn}
%\bibitem{Hosotani:1983vn}
%  Y.~Hosotani,
  %``Dynamical Gauge Symmetry Breaking As The Casimir Effect,''
  Phys.\ Lett.\  B {\bf 129} (1983) 193.
  %%CITATION = PHLTA,B129,193;%%

%\cite{Agashe:2004rs}
\bibitem{Agashe:2004rs}
  K.~Agashe, R.~Contino and A.~Pomarol,
  %``The minimal composite Higgs model,''
  Nucl.\ Phys.\  B {\bf 719} (2005) 165
  [arXiv:hep-ph/0412089].
  %%CITATION = NUPHA,B719,165;%%

%\cite{Agashe:2005dk}
\bibitem{Agashe:2005dk}
  K.~Agashe and R.~Contino,
  %``The minimal composite Higgs model and electroweak precision tests,''
  Nucl.\ Phys.\  B {\bf 742}, 59 (2006)
  [arXiv:hep-ph/0510164].
  %%CITATION = NUPHA,B742,59;%%
  
%\cite{Contino:2006qr}
\bibitem{Contino:2006qr}
  R.~Contino, L.~Da Rold and A.~Pomarol,
  %``Light custodians in natural composite Higgs models,''
  Phys.\ Rev.\  D {\bf 75} (2007) 055014
  [arXiv:hep-ph/0612048].
  %%CITATION = PHRVA,D75,055014;%%

%\cite{Panico:2008bx}
\bibitem{Panico:2008bx}
  G.~Panico, E.~Ponton, J.~Santiago and M.~Serone,
  %``Dark Matter and Electroweak Symmetry Breaking in Models with Warped Extra
  %Dimensions,''
  arXiv:0801.1645 [hep-ph].
  %%CITATION = ARXIV:0801.1645;%%

%\cite{Agashe:2003zs}
\bibitem{Agashe:2003zs}
  K.~Agashe, A.~Delgado, M.~J.~May and R.~Sundrum,
  %``RS1, custodial isospin and precision tests,''
  JHEP {\bf 0308} (2003) 050
  [arXiv:hep-ph/0308036].
  %%CITATION = JHEPA,0308,050;%%

%\cite{Agashe:2006at}
\bibitem{Agashe:2006at}
  K.~Agashe, R.~Contino, L.~Da Rold and A.~Pomarol,
  %``A custodial symmetry for Z b anti-b,''
  Phys.\ Lett.\  B {\bf 641} (2006) 62
  [arXiv:hep-ph/0605341].
  %%CITATION = PHLTA,B641,62;%%

%\cite{ArkaniHamed:2001nc}
\bibitem{ArkaniHamed:2001nc}
  N.~Arkani-Hamed, A.~G.~Cohen and H.~Georgi,
  %``Electroweak symmetry breaking from dimensional deconstruction,''
  Phys.\ Lett.\  B {\bf 513} (2001) 232
  [arXiv:hep-ph/0105239];
  %%CITATION = PHLTA,B513,232;%%
%\cite{ArkaniHamed:2002qy}
%\bibitem{ArkaniHamed:2002qy}
  N.~Arkani-Hamed, A.~G.~Cohen, E.~Katz and A.~E.~Nelson,
  %``The littlest Higgs,''
  JHEP {\bf 0207} (2002) 034
  [arXiv:hep-ph/0206021].
  %%CITATION = JHEPA,0207,034;%%

%\cite{Giudice:2007fh}
\bibitem{Giudice:2007fh}
  G.~F.~Giudice, C.~Grojean, A.~Pomarol and R.~Rattazzi,
  %``The Strongly-Interacting Light Higgs,''
  JHEP {\bf 0706} (2007) 045
  [arXiv:hep-ph/0703164].
  %%CITATION = JHEPA,0706,045;%%

%\cite{Barbieri:2007bh}
\bibitem{Barbieri:2007bh}
  R.~Barbieri, B.~Bellazzini, V.~S.~Rychkov and A.~Varagnolo,
  %``The Higgs boson from an extended symmetry,''
  Phys.\ Rev.\  D {\bf 76} (2007) 115008
  [arXiv:0706.0432 [hep-ph]].
  %%CITATION = PHRVA,D76,115008;%%

%\cite{Lodone:2008yy}
\bibitem{Lodone:2008yy}
  P.~Lodone,
  %``Vector-like quarks in a composite Higgs model,''
  arXiv:0806.1472 [hep-ph].
  %%CITATION = ARXIV:0806.1472;%%
%\cite{Gillioz:2008hs}
\bibitem{Gillioz:2008hs}
  M.~Gillioz,
  %``A light composite Higgs boson facing electroweak precision tests,''
  arXiv:0806.3450 [hep-ph].
  %%CITATION = ARXIV:0806.3450;%%

%\cite{Falkowski:2007hz}
\bibitem{Falkowski:2007hz}
  A.~Falkowski,
  %``Pseudo-Goldstone Higgs Production via Gluon Fusion,''
  Phys.\ Rev.\  D {\bf 77} (2008) 055018
  [arXiv:0711.0828 [hep-ph]].
  %%CITATION = PHRVA,D77,055018;%%

%\cite{Peskin:1991sw}
\bibitem{Peskin:1991sw}
  M.~E.~Peskin and T.~Takeuchi,
  %``Estimation of oblique electroweak corrections,''
  Phys.\ Rev.\  D {\bf 46} (1992) 381.
  %%CITATION = PHRVA,D46,381;%%

%\cite{Han:2004az}
\bibitem{Han:2004az}
  Z.~Han and W.~Skiba,
  %``Effective theory analysis of precision electroweak data,''
  Phys.\ Rev.\  D {\bf 71}, 075009 (2005)
  [arXiv:hep-ph/0412166];
  %%CITATION = PHRVA,D71,075009;%%
%\cite{Han:2005pr}
%\bibitem{Han:2005pr}
  Z.~Han,
  %``Electroweak constraints on effective theories with U(2) x U(1) flavor
  %symmetry,''
  Phys.\ Rev.\  D {\bf 73}, 015005 (2006)
  [arXiv:hep-ph/0510125].
  %%CITATION = PHRVA,D73,015005;%%

%\cite{Collaboration:2008ub}
\bibitem{Collaboration:2008ub}
  ALEPH~Collaboration {\it et al.},
  %``Precision Electroweak Measurements and Constraints on the
  %Standard Model,'' 
  arXiv:0811.4682 [hep-ex].
  %%CITATION = ARXIV:0811.4682;%%

%\cite{Barbieri:2008cc}
\bibitem{Barbieri:2008cc}
  R.~Barbieri, G.~Isidori, V.~S.~Rychkov and E.~Trincherini,
  %``Heavy Vectors in Higgs-less models,''
  arXiv:0806.1624 [hep-ph].
  %%CITATION = ARXIV:0806.1624;%%

%\cite{Carena:2006bn}
\bibitem{Carena:2006bn}
  M.~S.~Carena, E.~Ponton, J.~Santiago and C.~E.~M.~Wagner,
  %``Light Kaluza-Klein states in Randall-Sundrum models with custodial
  %SU(2),''
  Nucl.\ Phys.\  B {\bf 759} (2006) 202
  [arXiv:hep-ph/0607106];
  %%CITATION = NUPHA,B759,202;%%
%\cite{Carena:2007ua}
%\bibitem{Carena:2007ua}
%  M.~S.~Carena, E.~Ponton, J.~Santiago and C.~E.~M.~Wagner,
  %``Electroweak constraints on warped models with custodial symmetry,''
  Phys.\ Rev.\  D {\bf 76} (2007) 035006
  [arXiv:hep-ph/0701055].
  %%CITATION = PHRVA,D76,035006;%%

%\cite{Haisch:2007ia}
\bibitem{Haisch:2007ia}
  U.~Haisch and A.~Weiler,
  %``Determining the Sign of the Z-Penguin Amplitude,''
  Phys.\ Rev.\  D {\bf 76}, 074027 (2007)
  [arXiv:0706.2054 [hep-ph]].
  %%CITATION = PHRVA,D76,074027;%%

%\cite{Denner:1991kt}
\bibitem{Denner:1991kt}
  A.~Denner,
  %``Techniques for calculation of electroweak radiative corrections at the one
  %loop level and results for W physics at LEP-200,''
  Fortsch.\ Phys.\  {\bf 41}, 307 (1993)
  [arXiv:0709.1075 [hep-ph]].
  %%CITATION = FPYKA,41,307;%%

%\cite{Lavoura:1992np}
\bibitem{Lavoura:1992np}
  L.~Lavoura and J.~P.~Silva,
  %``The Oblique Corrections From Vector - Like Singlet And Doublet Quarks,''
  Phys.\ Rev.\  D {\bf 47}, 2046 (1993).
  %%CITATION = PHRVA,D47,2046;%%

%\cite{Davydychev:1991va}
\bibitem{Davydychev:1991va}
  A.~I.~Davydychev,
  %``A Simple formula for reducing Feynman diagrams to scalar integrals,''
  Phys.\ Lett.\  B {\bf 263}, 107 (1991).
  %%CITATION = PHLTA,B263,107;%%

%\cite{Anastasiou:1999bn}
\bibitem{Anastasiou:1999bn}
  C.~Anastasiou, E.~W.~N.~Glover and C.~Oleari,
  %``The two-loop scalar and tensor pentabox graph with light-like legs,''
  Nucl.\ Phys.\  B {\bf 575}, 416 (2000)
  [Erratum-ibid.\  B {\bf 585}, 763 (2000)]
  [arXiv:hep-ph/9912251].
  %%CITATION = NUPHA,B575,416;%%

%\cite{Anastasiou:2004vj}
\bibitem{Anastasiou:2004vj}
  C.~Anastasiou and A.~Lazopoulos,
  %``Automatic integral reduction for higher order perturbative  calculations,''
  JHEP {\bf 0407}, 046 (2004)
  [arXiv:hep-ph/0404258].
  %%CITATION = JHEPA,0407,046;%%

%\cite{Bamert:1996px}
\bibitem{Bamert:1996px}
  P.~Bamert, C.~P.~Burgess, J.~M.~Cline, D.~London and E.~Nardi,
  %``R_b and New Physics: A Comprehensive Analysis,''
  Phys.\ Rev.\  D {\bf 54}, 4275 (1996)
  [arXiv:hep-ph/9602438].
  %%CITATION = PHRVA,D54,4275;%%

%\cite{Bobeth:1999ww}
\bibitem{Bobeth:1999ww}
  C.~Bobeth, M.~Misiak and J.~Urban,
  %``Matching conditions for b --> s gamma and b --> s gluon in extensions  of
  %the standard model,''
  Nucl.\ Phys.\  B {\bf 567}, 153 (2000)
  [arXiv:hep-ph/9904413].
  %%CITATION = NUPHA,B567,153;%%

%\cite{Freitas:2008vh}
\bibitem{Freitas:2008vh}
  A.~Freitas and U.~Haisch,
  %``Anti-B --> X(s) gamma in two universal extra dimensions,''
  Phys.\ Rev.\  D {\bf 77}, 093008 (2008)
  [arXiv:0801.4346 [hep-ph]].
  %%CITATION = PHRVA,D77,093008;%%

%\cite{Barberio:2008fa}
\bibitem{Barberio:2008fa}
  E.~Barberio {\it et al.}  [Heavy Flavor Averaging Group],
  %``Averages of b-hadron and c-hadron Properties at the End of 2007,''
  arXiv:0808.1297 [hep-ex].
  %%CITATION = ARXIV:0808.1297;%%

%\cite{Hahn:2004fe}
\bibitem{Hahn:2004fe}
  T.~Hahn,
  %``CUBA: A library for multidimensional numerical integration,''
  Comput.\ Phys.\ Commun.\  {\bf 168}, 78 (2005)
  [arXiv:hep-ph/0404043].
  %%CITATION = CPHCB,168,78;%%

%\cite{Pomarol:2008bh}
\bibitem{Pomarol:2008bh}
  A.~Pomarol and J.~Serra,
  %``Top Quark Compositeness: Feasibility and Implications,''
  Phys.\ Rev.\  D {\bf 78}, 074026 (2008)
  [arXiv:0806.3247 [hep-ph]].
  %%CITATION = PHRVA,D78,074026;%%

%\cite{Abazov:2008sz}
\bibitem{Abazov:2008sz}
  V.~M.~Abazov {\it et al.}  [D0 Collaboration],
  %``Search for anomalous $\boldmath{Wtb}$ couplings in single top quark
  %production,''
  Phys.\ Rev.\ Lett.\  {\bf 101}, 221801 (2008)
  [arXiv:0807.1692 [hep-ex]];
  %%CITATION = PRLTA,101,221801;%%
%\cite{Aaltonen:2008sy}
%\bibitem{Aaltonen:2008sy}
  T.~Aaltonen {\it et al.}  [CDF Collaboration],
  %``Measurement of the Single Top Quark Production Cross Section at CDF,''
  arXiv:0809.2581 [hep-ex].
  %%CITATION = ARXIV:0809.2581;%%

\bibitem{JAAS:talk}
J.A. Aguilar-Saavedra, in 3rd Top Workshop at Grenoble: From Tevatron
to Atlas. 

%\cite{AguilarSaavedra:2008gt}
\bibitem{AguilarSaavedra:2008gt}
  J.~A.~Aguilar-Saavedra,
  %``Single top quark production at LHC with anomalous Wtb couplings,''
  Nucl.\ Phys.\  B {\bf 804}, 160 (2008)
  [arXiv:0803.3810 [hep-ph]].
  %%CITATION = NUPHA,B804,160;%%

%\cite{Baur:2005wi}
\bibitem{Baur:2005wi}
  U.~Baur, A.~Juste, D.~Rainwater and L.~H.~Orr,
  %``Improved measurement of t t Z couplings at the LHC,''
  Phys.\ Rev.\  D {\bf 73}, 034016 (2006)
  [arXiv:hep-ph/0512262].
  %%CITATION = PHRVA,D73,034016;%%

%\cite{Lazopoulos:2008de}
\bibitem{Lazopoulos:2008de}
  A.~Lazopoulos, T.~McElmurry, K.~Melnikov and F.~Petriello,
  %``Next-to-leading order QCD corrections to t-tbar-Z production at the LHC,''
  Phys.\ Lett.\  B {\bf 666}, 62 (2008)
  [arXiv:0804.2220 [hep-ph]].
  %%CITATION = PHLTA,B666,62;%%

%\cite{Cacciapaglia:2006gp}
\bibitem{Cacciapaglia:2006gp}
  G.~Cacciapaglia, C.~Csaki, G.~Marandella and J.~Terning,
  %``A New Custodian for a Realistic Higgsless Model,''
  Phys.\ Rev.\  D {\bf 75}, 015003 (2007)
  [arXiv:hep-ph/0607146];
  %%CITATION = PHRVA,D75,015003;%%
%\cite{Dennis:2007tv}
%\bibitem{Dennis:2007tv}
  C.~Dennis, M.~Karagoz Unel, G.~Servant and J.~Tseng,
  %``Multi-W events at LHC from a warped extra dimension with custodial
  %symmetry,''
  arXiv:hep-ph/0701158;
  %%CITATION = HEP-PH/0701158;%%
%\cite{Carena:2007tn}
%\bibitem{Carena:2007tn}
  M.~Carena, A.~D.~Medina, B.~Panes, N.~R.~Shah and C.~E.~M.~Wagner,
  %``Collider Phenomenology of Gauge-Higgs Unification Scenarios in Warped Extra
  %Dimensions,''
  Phys.\ Rev.\  D {\bf 77}, 076003 (2008)
  [arXiv:0712.0095 [hep-ph]].
  %%CITATION = PHRVA,D77,076003;%%

%\cite{Contino:2008hi}
\bibitem{Contino:2008hi}
  R.~Contino and G.~Servant,
  %``Discovering the top partners at the LHC using same-sign dilepton final
  %states,''
  JHEP {\bf 0806}, 026 (2008)
  [arXiv:0801.1679 [hep-ph]].
  %%CITATION = JHEPA,0806,026;%%

%\cite{Brooijmans:2008se}
\bibitem{Brooijmans:2008se}
  G.~H.~Brooijmans {\it et al.},
  %``New Physics at the LHC: A Les Houches Report. Physics at Tev Colliders 2007
  %-- New Physics Working Group,''
  arXiv:0802.3715 [hep-ph].
  %%CITATION = ARXIV:0802.3715;%%

%\cite{Atre:2008iu}
\bibitem{Atre:2008iu}
  A.~Atre, M.~Carena, T.~Han and J.~Santiago,
  %``Heavy Quarks Above the Top at the Tevatron,''
  arXiv:0806.3966 [hep-ph].
  %%CITATION = ARXIV:0806.3966;%%

%\cite{Contino:2004vy}
\bibitem{Contino:2004vy}
  R.~Contino and A.~Pomarol,
  %``Holography for fermions,''
  JHEP {\bf 0411}, 058 (2004)
  [arXiv:hep-th/0406257].
  %%CITATION = JHEPA,0411,058;%%

%\cite{Contino:2006nn}
\bibitem{Contino:2006nn}
  R.~Contino, T.~Kramer, M.~Son and R.~Sundrum,
  %``Warped/Composite Phenomenology Simplified,''
  JHEP {\bf 0705}, 074 (2007)
  [arXiv:hep-ph/0612180].
  %%CITATION = JHEPA,0705,074;%%

%\cite{Falkowski:2007iv}
\bibitem{Falkowski:2007iv}
 A.~Falkowski, S.~Pokorski and J.~P.~Roberts,
 %``Modelling strong interactions and longitudinally polarized vector boson
 %scattering,''
 JHEP {\bf 0712}, 063 (2007)
 [arXiv:0705.4653 [hep-ph]].
 %%CITATION = JHEPA,0712,063;%%

%\cite{Agashe:2006hk}
\bibitem{Agashe:2006hk}
  K.~Agashe, A.~Belyaev, T.~Krupovnickas, G.~Perez and J.~Virzi,
  %``LHC signals from warped extra dimensions,''
  Phys.\ Rev.\  D {\bf 77}, 015003 (2008)
  [arXiv:hep-ph/0612015];
  %%CITATION = PHRVA,D77,015003;%%
%\cite{Lillie:2007yh}
%\bibitem{Lillie:2007yh}
  B.~Lillie, L.~Randall and L.~T.~Wang,
  %``The Bulk RS KK-gluon at the LHC,''
  JHEP {\bf 0709}, 074 (2007)
  [arXiv:hep-ph/0701166];
  %%CITATION = JHEPA,0709,074;%%
%\cite{Agashe:2007ki}
%\bibitem{Agashe:2007ki}
  K.~Agashe {\it et al.},
  %``LHC Signals for Warped Electroweak Neutral Gauge Bosons,''
  Phys.\ Rev.\  D {\bf 76}, 115015 (2007)
  [arXiv:0709.0007 [hep-ph]];
  %%CITATION = PHRVA,D76,115015;%%
%\cite{Agashe:2008jb}
%\bibitem{Agashe:2008jb}
  K.~Agashe, S.~Gopalakrishna, T.~Han, G.~Y.~Huang and A.~Soni,
  %``LHC Signals for Warped Electroweak Charged Gauge Bosons,''
  arXiv:0810.1497 [hep-ph].
  %%CITATION = ARXIV:0810.1497;%%

\end{thebibliography}
\end{document}